\documentclass[letterpaper,twocolumn,prx,aps,showpacs,floatfix,superscriptaddress,nofootinbib]{revtex4-1}
\usepackage[latin1]{inputenc}
\usepackage{bm}
\usepackage{multirow}
\usepackage{amssymb}
\usepackage{amsbsy}
\usepackage{amsmath}
\usepackage{graphicx}
\usepackage{epsfig}
\usepackage{placeins}
\usepackage{MnSymbol}
\usepackage[usenames,dvipsnames]{color}
\usepackage[colorlinks,linkcolor=blue,citecolor=blue,urlcolor=blue,breaklinks=true]{hyperref}

\DeclareMathOperator{\Tr}{Tr}

\makeatletter

\begin{document} 
\title{Stiffness of Probability  Distributions of Work and Jarzynski Relation for Non-Gibbsian Initial States}

 \author{Daniel Schmidtke}
 \email{danischm@uos.de}
 \affiliation{Department of Physics, University of Osnabr\"uck, D-49069 Osnabr\"uck, Germany}
 
 \author{Lars Knipschild}
 \email{lknipschild@uos.de}
 \affiliation{Department of Physics, University of Osnabr\"uck, D-49069 Osnabr\"uck, Germany}
 
 \author{Michele Campisi}
 \affiliation{Dipartimento di Fisica e Astronomia,
Universit\`a di Firenze, via G. Sansone 1, I-50019 Sesto Fiorentino, Italy}
 
  \author{Robin Steinigeweg}
 \affiliation{Department of Physics, University of Osnabr\"uck, D-49069 Osnabr\"uck, Germany}
 
 \author{Jochen Gemmer}
 \email{jgemmer@uos.de}
 \affiliation{Department of Physics, University of Osnabr\"uck, D-49069 Osnabr\"uck, Germany}

\begin{abstract}
We consider closed quantum systems (into which baths may be integrated) that are driven, i.e., subject to time-dependent Hamiltonians. Our point of departure is 
the assumption that, if systems start in microcanonical states at some initial energies, the resulting probability distributions of work may be largely independent 
of the specific initial energies. It is demonstrated that this assumption has some far-reaching consequences, e.g., it implies the validity of the Jarzynski relation 
for a large class of non-Gibbsian initial states. By performing numerical analysis on integrable and non-integrable spin systems, we find the 
above assumption fulfilled for all considered examples. Through an analysis based on Fermi's Golden Rule, we partially relate these findings to the applicability of the 
eigenstate thermalization ansatz to the respective driving operators.
\end{abstract}



\maketitle

\section{Introduction}
The long-standing question regarding whether, and in which way, closed finite quantum systems approach thermal equilibrium has recently gathered renewed attention. On the theoretical side thermalization and equilibration have been investigated e.g.\ for rather abstract settings \cite{Popescu2006,Goldstein2006,Reimann2008,Reimann2015,Eisert2015,Gogolin2016} and also for more specific condensed-matter type systems \cite{Rigol2008,Steinigeweg2013,Beugeling2014,Steinigeweg2014}. In these works major concepts are the eigenstate thermalization hypothesis (ETH) and typicality, both of which will also play certain roles in the  paper at hand. The developments on experiments on ultra-cold atoms now allow for testing what have been merely theoretical results before; see e.g.\  Ref.\  \cite{Reimann2016, balz17,tomkovic}.\\
Rather than just the existence of equilibration within closed quantum systems, lately  the very peculiarities of the  dynamical approach to equilibrium have moved to the center of interest \cite{Malabarba2014,Reimann2016}. 
Questions addressed in this context include limits on relaxation time scales and agreement of unitary quantum dynamics of closed quantum systems with standard statistical relaxation principles, such as Fokker-Planck equations \cite{thikonenkov13, ates12,Niemeyer2013,Niemeyer2014}, or more general, standard stochastic processes \cite{Gemmer2014,Schmidtke2016}. But also the emergence of universal non-equilibrium behavior involving work and driven systems is under  discussion at present \cite{Miller2017}.

To a large extent universal non-equilibrium behavior may be captured by fluctuation theorems, see e.g.\  Ref.~\cite{Seifert2008} and references therein. The Jarzynski relation (JR), a general statement on work that has to be invested to drive processes also and especially far from equilibrium, is a prime example of such a fluctuation theorem. Many derivations of the JR from various starting grounds have been presented. These include classical Hamiltonian dynamics, stochastic dynamics such as Langevin or master equations, as well as quantum mechanical
starting points \cite{Seifert2008,Roncaglia2014, Haenggi2015, Seifert2008,Esposito2009,Jarzynski2011,cuendet06}. However, all these derivations assume that the system, that is acted on with some kind of ``force'', is strictly in a Gibbsian equilibrium state before the process starts. (The notion of ``the system'' here routinely includes the bath.) Thus, this starting point differs significantly from the progresses in the field of thermalization: There, the general features of thermodynamic relaxation are found to emerge entirely from the system itself without any necessity of evoking external baths or specifying initial states in detail. Clearly, the preparation of a strictly Gibbsian initial state requires the coupling to a bath prior to starting the process.

This situation renders  the question whether or not the JR is valid for  systems starting in other states than Gibbsian states rather exigent. Clarifying this question is the main purpose of the paper at hand. Since counterexamples may  be constructed, there cannot be any affirmative answer without restrictions on the quantum system and the process protocol. However, previous works \cite{Talkner2008, Campisi2008, Talkner2013, Campisi2011} have shown that, when the initial state is microcanonical, the JR does not follow, but a related entropy-from-work relation emerges instead. The question remains, however, if and under what conditions the JR holds approximately for non-canonical initial states. Thus, the emphasis in the search for the origins of the JR's validity is shifted from specifying the initial state to specifying the nature of the system.

An intimately related question has lately also been addressed in Ref.~\cite{sagawa}.~There the validity of fluctuation theorems for typical pure initial states is traced back to certain features of the bath-system, including the existence of a ``quantum speed limit'' (Lieb-Robinson bounds) within the bath. While this argument is  certainly valid, it requires rather large baths to apply for enduring processes. In the examples discussed below, processes are too long and systems are too small to employ the argument given in Ref.\  \cite{sagawa} to explain the validity of the JR. Indeed the paper at hand offers a largely unrelated, alternative approach.

The paper at hand is organized as follows: Sec. \ref{sec:workdistri} starts with the introduction of our basic hypothesis of probability distribution functions of work (work pdf's) being largely independent of the respective energy for microcanonical initial states. The validity of the JR is shown to follow from this assumption. In order to quantify this independence (``stiffness''), we discuss an appropriate measure which is one main target of the later numerical investigations.  In Sec. \ref{sec:model} we describe our spin models as well as the specific microcanonical initial states and the general form of the  work-inducing protocols. In Sec. \ref{workdistri} and \ref{sec:prechain} we numerically analyze the dynamics of our models without any driving, in order to identify interesting specific parameters for the driving protocols. After these introductive sections we finally present in Sec. \ref{sec:work_spin_ladder} and \ref{sec:work_spin_chain} the numerical results on the stiffness of the work pdf's. In order to provide indications that our results may be understood by more fundamental concepts, i.e., Fermi's Golden Rule (FGR) and eigenstate thermalization hypothesis,  we also present matrix representations of the driving operators and discuss them in Sec. \ref{fgr}. At last we briefly summarize in Sec. \ref{conclusion} and draw conclusions.


\section{Stiffness of Work Pdf's and Jarzynski Relation for Non-Gibbsian Initial States}\label{sec:workdistri}

The analysis at hand focuses exclusively on closed systems. While it is physically appropriate to interpret the examples in Sec. \ref{sec:model} in terms of ``considered system'' and ``environment'' or ``bath'', we technically treat the system+environment compound regardless of the coupling strength as one closed system. Thus, since there is no external source or sink of heat, any energy change of the full system is to be counted as work $W$ (for an overview over different perspectives, see e.g.\  Ref.\  \cite{ueberblickJanet}.) In this respect we choose the same starting point as employed in derivations of the JR as described, e.g.\ , in Ref.\  \cite{Campisi2011} and references therein. However, while in Ref. \cite{Campisi2011} the
assumption of a canonical, Gibbsian initial state is of vital importance, we base our consideration on much larger classes of initial states of the full system. The central role which the assumption of strictly Gibbsian state plays in the afore mentioned works is replaced by the assumption of ``stiffness'' of the work-pdf's (as introduced in detail below in Eq.~\ref{eq:stiff}) 

\subsection{Stiffness of Work pdf's and Jarzynski Relation for Microcanonical Initial States}
We now embark on the detailed presentation of our approach to a derivation of the JR for microcanonical initial states. Our initial states are given in terms of projection operators $\pi^\alpha_{E,\delta }$. These $\pi^\alpha_{E,\delta }$ are spanned by the energy eigenstates of some Hamilton operator $H_{\alpha}$, featuring eigenvalues from the interval $[E-\delta/2, E+\delta/2]$. The parameter $\alpha$ may take on the values $i,f$, indicating  ``initial'' and ``final'', respectively. The microcanonical initial states $\rho_i(E)$ may then be simply written as:
\begin{equation}
 \label{eq:inistat}
 \rho_i(E):= \pi^i_{E,\delta} / \text{Tr} \{ \pi^i_{E,\delta} \}~~~.
\end{equation}
The considered process is driven by a time-dependent Hamiltonian $H(t)$. We denote the duration of the driving by $T$ such that $H_i = H(0)$ and $H_f = H(T)$. For later reference we also introduce the Hamiltonian $\tilde H(t)$ which implements the ``backward protocol'' as $\tilde H(t):= H(T-t)$. The forward protocol induces a unitary time-propagation operator ${\bf U}(t)$ defined by:
\begin{equation}
 {\bf U}(t) := \mathcal{T} \exp\left(-i\,\int^t_0 H(t') dt'\right) ~~~,
 \label{eq:U}
\end{equation}
where $\mathcal{T}$ is the time-ordering operator and we tacitly set $\hbar =1$. The time propagation for the backward protocol $\tilde{{\bf U}}(t)$ is defined by an completely 
analogous expression based on $\tilde{H}(t)$. Now we are set to define the work pdf for the forward protocol $p_E(W)$, namely the probability density with which the process consumes an amount of work $W$ if the system was initialized at energy $E$ and driven according to $H(t)$:
\begin{equation}
 p_E(W) := \frac{1}{\delta} \text{Tr}\{ \pi^f_{E+W,\delta}\, {\bf U}(T)\, \rho_i(E)\,{\bf U}^\dagger(T) \} 
 \label{eq:PEW}
\end{equation}
Here, $\delta$ is to be chosen large compared to the level spacing of the full system, but small compared to the involved energy scales of $E, W$. We assume that a range of such $\delta$ exists for which $p_E(W)$ indeed becomes approximately independent of $\delta$. (All our numerical investigations at hand agree with this assumption, see Appendix \ref{A-chi_sec}). Obviously,  $p_E(W)$ may equally well be simply perceived as the transition probability density with which the full system ends up at energy $E+W$ if it started at $E$ under the forward driving. By $\tilde{p}_{E'}(W')$ we denote the corresponding work pdf for the backward protocol. Its definition is completely analogous to Eq.\ (\ref{eq:PEW}), but  with the replacements $E \rightarrow E', W \rightarrow W',  {\bf U} (T) \rightarrow \tilde{{\bf U}}(T), i \leftrightarrow f$.  {We remark already here that below in Eq.\ (\ref{eq:stiff}) we will assume  a certain property of the work pdf's $p_E(W),\,\tilde{p}_{E'}(W')$. While this assumption appears physically plausible, we want to emphasize here that acquiring any information on the work pdf's requires the solution of the time-dependent Schr\"odinger equation for the full system, since Eq.\ (\ref{eq:PEW}) comprises the time propagation operator  ${\bf U}(t)$. Computing the latter is in general accompanied by great numerical efforts, especially when the Hamiltonian is explicitly time-dependent and the underlying Hilbert space is high-dimensional, as is usually the case for relevant setups in this context; see e.g.\  Refs. \cite{deraedt1987,Jin2016} and Appendix \ref{A-tdse}.}

However, irrespective of the concrete forms of the work pdf's, microcanonical fluctuation theorems \cite{Talkner2007,Talkner2008} offer a connection between the work pdf's and the densities of states (DOS) $\Omega_\alpha$  of the corresponding Hamiltonians  $H_\alpha$:
\begin{equation}
 \frac{p_E(W)}{\tilde{p}_{E+W}(-W)} = \frac{\Omega_f(E+W)}{\Omega_i(E)}~~~~.
 \label{eq:mcft}
\end{equation}
This microcanonical fluctuation theorem is expected to hold for dynamics as generated by microreversible Hamiltonians $H(t)$ \cite{Talkner2013}. We note here that the models analyzed below are not microreversible in the standard sense \cite{Talkner2013,sakurai}, since they involve magnetic fields. However, due to a property which is somewhat similar to microreversibility, Eq.\ (\ref{eq:mcft}) applies to these models as well; for details see Appendix \ref{A-deri4}.

Next we define a property of the work pdf's which we call ``stiffness''. We call a forward work pdf $p_E(W)$ stiff if does not depend on the  initial energy $E$. Likewise we call a backward work pdf $\tilde{p}_{E+W}(-W)$ stiff if it does not depend on its initial energy $E+W$. To be more explicit, stiffness implies
\begin{equation}
p_E(W) = p_{E_0}(W) \quad ,~~   \tilde{p}_{E+W}(-W) = \tilde{p}_{E_0}(-W)~~~,\label{eq:stiff}
\end{equation}
where $E_0$ is some fixed, constant energy. Note that Eq.\ (\ref{eq:stiff}) is the central assumption from which the validity of the JR for microcanonical initial states will eventually be inferred, as announced above. Of course, Eq.\ (\ref{eq:stiff}) is not expected to hold for all energies $E$. Here we only require that the relations in Eq.\ (\ref{eq:stiff}) hold for energies $E$ from an energy interval $[E_0 - \Delta/2, E_0 +\Delta/2]$ such that
\begin{equation}
 \label{eq:int}
\int_{E_0 - \Delta/2}^{E_0 +\Delta/2} p_{E_0}(W)\, dW \approx 1.
\end{equation}
This means, the stiffness Eq.\ (\ref{eq:stiff}) must hold at least for an energy interval which is large enough to comprise almost the entire work pdf $p_{E_0}(W)$.

If the work pdf's are indeed stiff, Eq.\ (\ref{eq:mcft}) may be rewritten as
\begin{equation}
 \frac{p_{E_0}(W)}{\tilde{p}_{E_0}(-W)} = \frac{\Omega_f(E+W)}{\Omega_i(E)}~~~~.
 \label{eq:mcft1}
\end{equation}
Obviously, the l.h.s. of Eq.\ (\ref{eq:mcft1}) no longer depends on energy $E$. This, however, restricts the possible functional forms of $ \Omega_i,  \Omega_f$ to the following:
\begin{equation}
 \label{eq:mustexp}
 \Omega_i(U) \stackrel{!}{=} Z_i e^{\beta U}, \quad   \Omega_f(U) \stackrel{!}{=} Z_f e^{\beta U}~~~,
\end{equation} 
where $U$ parametrizes the inner energy and should not be confused with the previously introduced time propagation operator. (So far, $Z_i,Z_f,\beta$ are just constants. Later on they will take the roles of partition functions and inverse temperature, respectively.) This implies that stiff work pdf's can only occur if the DOS's of the initial and the final Hamiltonian are both exponential in energy at least on the energy interval where the pdf's are non negligible, with the same energy prefactor, i.e.,  $\beta$. Note,  however, that the converse does not hold, i.e., DOS's in agreement with Eq.\ (\ref{eq:mustexp}) do by themselves neither imply stiffness of the work pdf's, nor the validity of the JR for microcanonical initial states. Plugging Eq.\ (\ref{eq:mustexp}) back into Eq.\ (\ref{eq:mcft}) and rearranging a little yields:
\begin{equation}
 \label{eq:foba}
  \tilde{p}_{E+W}(-W) = p_{E}(W) \frac{Z_i}{Z_f} e^{- \beta W}~~~. 
\end{equation}
In general, the backward work pdf's do not necessarily sum up to unity in the following sense \cite{Campisi2008}: $\int  \tilde{p}_{E+W}( W) dW \neq1$. However, under the assumption of stiffness of the backward work pdf's in Eq.\ (\ref{eq:stiff}), they do, i.e., 
\begin{equation}
 \label{eq:sumone}
 \int  \tilde{p}_{E+W}(-W) dW =  \int \, \tilde{p}_{E_0}(-W) dW  = 1~~~.
\end{equation}
Upon inserting Eq.\ (\ref{eq:foba}) into Eq.\ (\ref{eq:sumone}) and rearranging one obtains
\begin{equation}
 \label{eq:jarmic}
 \int p_{E}(W)e^{- \beta W} dW =  \frac{Z_f}{Z_i}~~~,
\end{equation}
which formally is a JR for the work pdf's obtained by starting from microcanonical initial states, with the temperature replaced by a parameter describing the exponential growth of the DOS of the full system. As such Eq.\ (\ref{eq:jarmic}) already represents the main result of the present section. Note that Eq.\ (\ref{eq:jarmic}) holds for arbitrary processes and its r.h.s. only contains static, process-independent model parameters. However, in order to demonstrate even closer analogy with the standard JR, it remains to be explained in which sense the r.h.s of Eq.\ (\ref{eq:jarmic}) may be  considered as the familiar r.h.s of the standard JR, $e^{-\beta \Delta F}$, where $F$ is the free energy. Such an identification would hold if 
\begin{equation}
 \label{eq:freeen}
 -\frac{\ln Z_\alpha}{\beta} \stackrel{?}{=} F_\alpha~~~.
\end{equation}
In order to judge whether or not Eq.\ (\ref{eq:freeen}) is justified, consider the logarithm of Eq.\ (\ref{eq:mustexp}),
\begin{equation}
 \label{eq:mustexpln}
 \ln  \Omega_\alpha =  \ln  Z_\alpha + \beta U~~~.
\end{equation}
If one identifies, along the lines of Boltzmann's original approach, the entropy $S_\alpha$ as
\begin{equation}
 \label{eq:boltzent}
 \ln  \Omega_\alpha := S_\alpha
 \end{equation}
(where we tacitly set $k_B =1$), one may convert Eq.\ (\ref{eq:mustexpln}) into
\begin{equation}
 \label{eq:freeident}
 -\frac{\ln Z_\alpha}{\beta} = U - \frac{S_\alpha}{\beta}.
\end{equation}

Note that, in accordance with Eq.~(\ref{eq:mustexp}), $\partial_U S_\alpha = \beta$, hence $\beta$ has the meaning of inverse temperature, and the r.h.s. of Eq.\ (\ref{eq:freeident}) is, accordingly the free energy $F$ as introduced in standard textbooks on phenomenological thermodynamics. In this sense  
Eq.\ (\ref{eq:freeen}) indeed holds, which entails the rewriting of Eq.\ (\ref{eq:jarmic}) in a form closer to the familiar one:
\begin{equation}
 \langle  e^{- \beta W} \rangle_E =e^{-\beta \Delta F}~~~,
\end{equation}
where $\langle \cdots \rangle_E$ denotes the microcanonical expectation value corresponding to energy $E$. This concludes our consideration on the validity of a JR for microcanonical initial states under the assumption of stiff work-pdfs. 

While we argue below that the JR
holds for an even larger class of initial states, the following Sects. may be skipped and reading continued at Sect. \ref{qstiff} for a first overview.

\subsection{Typical Validity of Jarzynski Relation for Random Pure States from an Energy Shell} \label{pure}

So far, only microcanonical initial states that are diagonal in the energy eigenbasis of the initial Hamiltonian have been considered. However, some arguments related to ``typicality'' suffice to establish that, given the validity of Eq.\ (\ref{eq:stiff}), the validity of a JR will hold, even for a very large majority of pure states. Consider to this end pure states $|\psi_{E,\delta} \rangle$ which are drawn at random according to the unitary invariant Haar measure from the Hilbert space spanned by the projector $\pi^i_{E,\delta}$. The corresponding work pdf is then given by
\begin{equation}
  p_E(W) := \frac{1}{\delta} \langle \psi_{E,\delta}| {\bf U}^\dagger(T)  \pi^f_{E+W,\delta}\, {\bf U}(T) |\psi_{E,\delta} \rangle \, .
\end{equation}
Of course, here $p_E(W)$ technically depends on the specific  $|\psi_{E,\delta} \rangle$. However, employing the methods and results of ``typicality'' \cite{qtbook, bartsch2009} one finds for the ``Hilbert-space average'' (HA$[ \ldots ]$) of  $p_E(W)$ over the above $|\psi_{E,\delta} \rangle$
\begin{equation}
 \mbox{HA}[ p_E(W)] = \frac{1}{\delta} \text{Tr}\{ \pi^f_{E+W,\delta}\, {\bf U}(T)\, \rho_i(E)\,{\bf U}^\dagger(T) \} \, , 
\end{equation}
which equals the corresponding result for the mixed, microcanonical initial state $\rho_i(E)$, cf.\ Eq.\ (\ref{eq:PEW}). While this finding points in the direction of the JR being fulfilled for the vast majority of the  $|\psi_{E,\delta} \rangle$, it is, by itself, not sufficient to conclude for the latter. In order to do so, it remains to be shown that the corresponding ```Hilbert-space variances'' (HV$[ \ldots ]$) is small. Fortunately, expressions for such Hilbertspace variances may also be found in the literature \cite{qtbook, bartsch2009}. Prior to computing these expressions for the current case we introduce some convenient notation. Let $\sigma^2(A)$ denote the variance of the spectrum of some operator $A$, with $A$ being Hermitian, i.e., featuring real eigenvalues. Then the Hilbert-space variance for the work pdf is given by 
 \cite{qtbook, bartsch2009}:
\begin{equation}
  \mbox{HV}[ p_E(W)]= \frac{\sigma^2( \pi^i_{E,\delta}   {\bf U}^\dagger(T)  \pi^f_{E+W,\delta}\, {\bf U}(T)  \pi^i_{E,\delta} )}{\mbox{Tr}\{ \pi^i_{E,\delta} \} +1}
\end{equation}
Since the operator for which the spectral variance has to be determined contains only projectors and unitaries, i.e., has only eigenvalues between zero and one, 
an upper bound on the spectral variance is readily  found:
\begin{equation}
 \sigma^2( \pi^i_{E,\delta}   {\bf U}^\dagger(T)  \pi^f_{E+W,\delta}\, {\bf U}(T)  \pi^i_{E,\delta} ) < 1
\end{equation}
(This bound may easily be tightened, but this is of no further relevance here). This yields an upper bound for the Hilbert-space variance
\begin{equation}
  \mbox{HV}[ p_E(W)]< \frac{1}{\mbox{Tr}\{ \pi^i_{E,\delta} \} +1} \, .
\end{equation}
The crucial quantity here is obviously $\mbox{Tr}\{ \pi^i_{E,\delta} \}$ which is just the number of eigenstates of the initial Hamiltonian $H_i$ within the energy interval of size $\delta$ around $E$. For any given  $\delta$ it is to be expected that this number of eigenstates increases quickly (exponentially) with increasing bath size. Hence, for large baths $ \mbox{HV}[ p_E(W)]$ becomes very small, thus rendering the outcome for $p_E(W)$ for the overwhelming majority of individual $|\psi_{E,\delta} \rangle$'s indeed very close to the outcome one obtains from the  microcanonical initial state $\rho_i(E)$. Or, to rephrase, all above findings on microcanonical initial states $\rho_i(E)$ transfer to pure initial states $|\psi_{E,\delta} \rangle$ for all practical purposes. In this sense the JR also applies to very many pure states. Moreover, this principle underlies all numerical calculations of work pdf's for larger systems presented in the paper at hand. Time evolutions of a mixed states $\rho_i(E)$ are simply replaced by time evolution of a randomly drawn, pure  $|\psi_{E,\delta} \rangle$, since the latter are numerically much less costly, cf. \ref{A-workpdf}

\subsection{Energetically Broader Initial States}
Next we discuss the generalization of the validity of the JR for initial states which are restricted to  an energy shell,  to  initial states that live within the region to which the stiffness assumption Eq.~(\ref{eq:stiff}) applies. For conciseness we only discuss generalized diagonal (w.r.t. $H_i$) initial states. However, the generalization to corresponding typical pure states analogous to the considerations in Sec.~\ref{pure} is straightforward. Such a more general initial state $R_i$ may be written as
\begin{equation}
 \label{genini}
 R_i := \sum_n K(E_n) \rho_i(E_n), \quad \sum_n K(E_n)=1~~~,
\end{equation}
where the $K(E_n)$ are the probabilities to find the system initially in the energy interval around $E_n$. As mentioned above, the work pdf's $p_E(W)$ may be viewed as conditional probability densities. Thus, the total probability density $p(W)$ to invest the work $W$ is given by
\begin{equation}
 \label{totprob}
 p(W) = \sum_n K(E_n)p_{E_n}(W)~~~.
\end{equation}
Computing the average exponential of work thus yields
\begin{eqnarray}
\label{jargen}
 \int p(W) e^{- \beta W} dW & = & \sum_n K(E_n)\int p_{E_n}(W)e^{- \beta W} dW \nonumber \\
 =  \sum_n K(E_n) \frac{Z_f}{Z_i} & = & \frac{Z_f}{Z_i}~~~,
 \end{eqnarray}
where we used Eq.\ (\ref{totprob}), (\ref{genini}) and (\ref{eq:jarmic}). This concludes the reasoning for the above validity of the JR for more general, diagonal initial states.

\subsection{Quantifying Stiffness} \label{qstiff}
Prior to investigating specific models and protocols in Sec.\ \ref{sec:model}, we now introduce our measure that allows us to quantify the ``quality`` of stiffness for specific cases in Sec. \ref{sec:work_spin_ladder} and \ref{sec:work_spin_chain}).
By making use of well-tailored numerical techniques we are able to compute the work pdf's $p_E(W), \tilde{p}_{E+W}(-W)$ at reasonable expense for system sizes that allow for valuable results; see below for details. (For conciseness we formally introduce the ``stiffness quantifier'' $\bar{\chi}$ for the forward processes only.) The difference for two work pdf's $p_{E}(W),\, p_{E^\prime}(W)$ induced by the same protocol, but for two different microcanonical initial states may most simply be quantified as 
\begin{equation}
\chi(E,E^\prime) := \int \left[p_{E^\prime}(W) - p_{E}(W)\right]^2 dW~~~.
 \label{eq:chi}
\end{equation}
Evidently, the  $\chi(E,E^\prime)$ are positive by construction. The mean difference of the work pdf's as resulting from microcanonical initial states from an energy region of size $\Delta$ around a central energy $E_0$ (cf.\  paragraph above Eq.\ (\ref{eq:int})~) from the work pdf corresponding to the initial state living at $E_0$ may be written as 
\begin{equation}
 \bar{\chi}:= \frac{1}{\Delta}\int_{E_0-\Delta/2}^{E_0-\Delta/2}    \chi(E_0, E^\prime) dE^\prime~~~.
 \label{eq:barchi}
\end{equation}
Since all  $\chi(E,E^\prime)$ are positive, a small $\bar{\chi}$  not only implies a small average, but also that most of the respective $\chi(E,E^\prime)$ must be individually small. Or, to rephrase, if one had $\bar{\chi}=0$, this would be sufficient to infer that $ \chi(E_0, E^\prime)=0$. Thus below we employ $\bar{\chi}$ as a reliable stiffness quantifier or, more specific, $\bar{\chi} \ll 1$ signals stiff work pdf's. 


\section{Models, initial states, and driving protocol} \label{sec:model}

\begin{figure}[t]
\includegraphics[width=0.7\columnwidth]{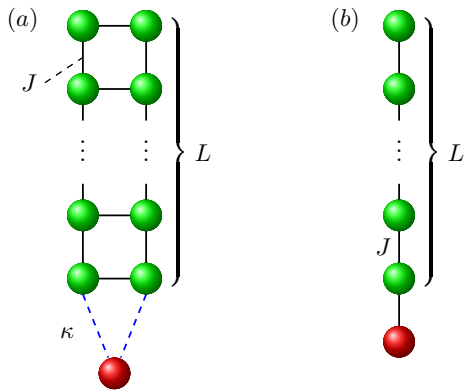}
\caption{Sketches of the spin models investigated here. (a) displays a spin ladder (green dots) acting as bath for an additional spin (red dot) which is coupled to the ladder with strength $\kappa$. Each leg of the ladder consists of $L$ spins. (b) displays an isotropic spin chain (green dots) also consisting of $L$ spins. Here the additional spin just enlarges the chain by one site.}
\label{fig:figure1}
\end{figure}

 {In the paper at hand, we focus our investigation on a model that can be associated to actual spin resonance experiments (see e.g.\ \cite{esr2,esr1}) since there driving protocols are easily implemented and one can without much efforts control essential features as e.g.\  the width of the distributions.}\\ 
For this purpose we consider an isotropic Heisenberg spin--1/2 ladder (denoted as bath) to which we attach an additional spin--1/2 as displayed in Fig.\ \ref{fig:figure1}(a). Non-integrable spin ladders (in the sense of the Bethe ansatz) and derivates of it are already intensively studied regarding e.g.\  relaxation of magnetization \cite{Niemeyer2013,Niemeyer2014,Schmidtke2016} or energy imbalances \cite{Steinigeweg2015,Khodja2016} and thus are ideal systems to start with. Note that, in order to allow for a non-trivial resonant driving protocol, we apply also a static magnetic field on the additional spin (see for details below).\\ Furthermore, we investigate a second model, which stands in contrast to the first, where we drop (i) the non-integrability, (ii) the ``distinction`` between bath and system, and (iii) the resonant driving. This results here in a simple spin--1/2 chain as displayed in Fig. \ref{fig:figure1}(b). Note that this chain model serves merely as comparator system and will not be investigated as thorough as the ladder model. \\
In both cases we allow only for next neighbor interaction. Thus the bath Hamiltonian for the ladder model reads
\begin{equation}
\begin{split}
H^\prime_{\text{ladder}} &= J \sum^2_{r=1} \sum^{L-1}_{i=1} \left(S^x_{i,r}S^x_{i+1,r}+S^y_{i,r}S^y_{i+1,r}+S^z_{i,r}S^z_{i+1,r}\right) \\
& + J \sum^{L}_{i=1} \left(S^x_{i,1}S^x_{i,2}+S^y_{i,1}S^y_{i,2}+S^z_{i,1}S^z_{i,2}\right)~~~,
 \end{split}
 \label{eq:Hamilton_ladder}
\end{equation}
where $S^{x,y,z}_{i,r}$ are spin--1/2 operators at site $(i,r)$. Likewise, the bath Hamiltonian for the chain model reads
\begin{equation}
H^\prime_{\text{chain}} = J \sum^{L-1}_{i=1}\left( S^x_{i}S^x_{i+1}+S^y_{i}S^y_{i+1}+S^z_{i}S^z_{i+1}\right)~~~. 
 \label{eq:Hamilton_chain}
\end{equation}
In all cases, $J$ denotes the exchange coupling constant which throughout this work is set to $J=1$.\\ 
As seen in Fig.\ \ref{fig:figure1}, for the ladder model we attach the additional spin in such a way that it only interacts with the last spin of each leg with coupling strength $\kappa$, whereas for the chain model the system spin simply enlarges the chain by one site. Thus, for the ladder model the bath-system interaction Hamiltonian reads
\begin{multline}
H^{\prime\prime}_{\text{ladder}} = J \left[ (S^x_{L,1}+S^x_{L,2}) S^x_{\text{sys}}+(S^y_{L,1}+S^y_{L,2})S^y_{\text{sys}}  \right.\\
\left.+(S^z_{L,1}+S^z_{L,2})S^z_{\text{sys}} \right]~~~.
 \label{eq:Hamilton_ladder2}
\end{multline}
where $S^{x,y,z}_{\text{sys}}$ denote the respective spin--1/2 operators of the additional (system) spin. Furthermore, we apply a static magnetic field $B$ in $z$ direction onto the additional spin, i.e.,
\begin{equation}
H_{\text{mag}} =  B S^z_{\text{sys}}~~~.
 \label{eq:Hamilton_B}
\end{equation}
Finally, the whole time-independent Hamiltonian (at time $t=0$) for the spin ladder model is given by 
\begin{equation}
H^{0}_{\text{ladder}} := H^\prime_{\text{ladder}} + \kappa H^{\prime\prime}_{\text{ladder}} + H_{\text{mag}}~~~.
\label{eq:full_Hamilton}
\end{equation}
In case of the chain model the bath-system interaction Hamiltonian reads
\begin{equation}
H^{\prime\prime}_{\text{chain}} = J \left[ S^x_{L} S^x_{\text{sys}}+S^y_{L} S^y_{\text{sys}} + S^z_{L}S^z_{\text{sys}} \right]
 \label{eq:Hamilton_chain2}
\end{equation}
and hence the whole time-independent Hamiltonian for the chain model is
\begin{equation}
H^{0}_{\text{chain}}  := H^\prime_{\text{chain}}  + H^{\prime\prime}_{\text{chain}} ~~~.
\label{eq:full_Hamilton_chain}
\end{equation}
Note that here no variable bath-system coupling parameter $\kappa$ appears since we set throughout this work for the chain model $\kappa=1$.\\
Next we explain the driving protocol which is here a time-dependent magnetic field in $x$ direction, applied onto the additional spin, similar to spin-resonance experiments; see e.g.\  \cite{esr2,esr1}. The time-dependent driving Hamiltonian reads
\begin{equation}
 H_D(t,\nu) = \lambda \sin(\nu t) S^x_{\text{sys}}~~~,
 \label{eq:driving}
\end{equation}
where $\lambda$ denotes the strength of the irradiation and $\nu$ its frequency.  {We remark here that for the ladder model we choose $\nu=B$ so that the driving is indeed resonant whereas for the chain model where no static magnetic field is applied the protocol is inevitably non-resonant.}
For convenience we will consider hereafter only time-reversal symmetric protocols, i.e., protocol durations with half integer number of periods since then $H(t)=\tilde{H}(t)$ and especially ${\bf U}(t)=\tilde{{\bf U}}(t)$ hold; cf.\  Appendix \ref{A-deri4}.

Finally we can state the time-dependent Hamiltonians describing the dynamics during the entire protocol as
\begin{equation}
 H_{l/c}(t) = H^0_{l/c}\;+\;H_D(t,\nu)~~~,
 \label{eq:ham_time}
\end{equation}
where the subscript denotes either the ladder model ($l$) or the chain model ($c$).  {At last we repeat as a reminder that our initial states are microcanonical states with energy width $\delta$.}
Here we use $\delta\approx0.07$ for all calculations; see for details Appendix \ref{A-initialst}. Since, however, generating such states requires exact diagonalization, which is only feasible for small systems, we employ for systems consisting of more than $15$ spins an approximation scheme based on typicality that allows for energetically sharp states; cf.\  \cite{Elsayed2013,Khodja2016,Jin2016}. Likewise, work distributions for large systems cannot be gained directly, i.e., we again need to employ an approximation scheme as discussed in Appendix \ref{A-workpdf}.\\

\begin{figure}
\includegraphics[width=1\columnwidth]{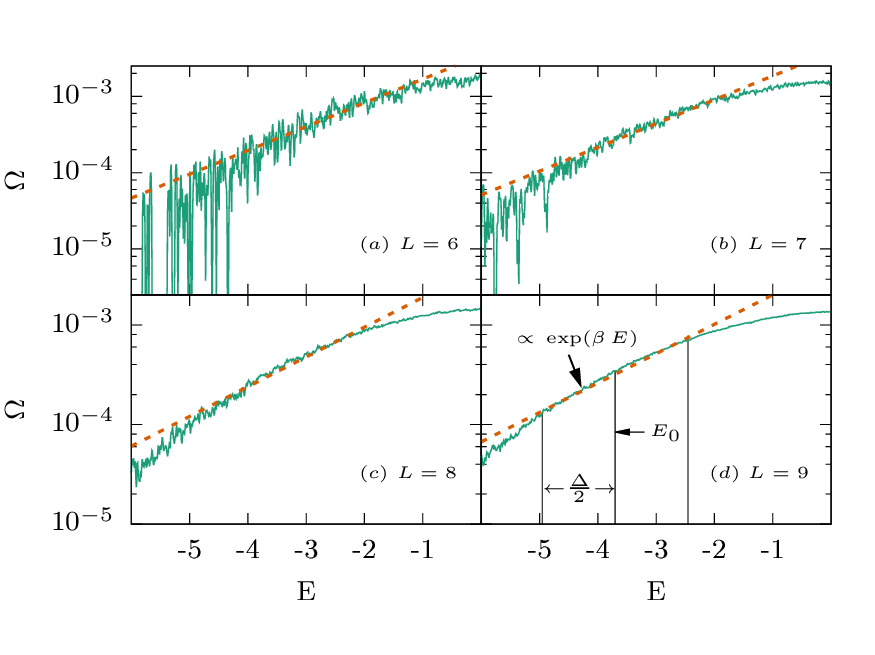}
\caption{Semi-log plots of the DOS for the ladder model for $L=6-9$ and $\kappa = 0.2$. Dashed lines indicate exponential fits to an intermediate energy regime of the displayed energy range.  {One clearly recognizes that $\Omega$ becomes smoother with increasing system size, as expected. The energy range where $\Omega$ agrees with exponential growth behavior seems to increase also.} Note that for $\kappa=0.6$ we find a very similar behavior.}
\label{fig:figure2}
\end{figure}


\section{preliminary considerations on the spin-ladder setup: density of states and free relaxation dynamics}\label{workdistri}

As argued in Sec. \ref{sec:workdistri}, the stiffness of the  work pdf's is conditioned on the DOS being exponential. Thus, prior to examining the work pdf's as resulting from an actual driving, we numerically analyze the DOS of the non-driven model in order to identify exponential regions. Later on we will choose our starting energies and  driving protocols such that the  work pdf's essentially remain within those exponential regions. Generally  we expect these exponential regions to become arbitrarily large with increasing system sizes.

Furthermore, physically meaningful interaction strengths $\kappa$, cf.~Eq.~(\ref{eq:full_Hamilton}), and  reasonable driving strengths $\lambda$, cf.~Eq.~(\ref{eq:driving}), have to be chosen, before an actual driving can be numerically performed. To this end  we will first analyze the relaxation dynamics of the magnetization of the additional spin (without any driving) and its dependence on $\kappa$. Computation of this relaxation behavior allows to distinguish between ``weak coupling'' (exponential decay) and ``strong coupling'' (non-exponential decay). We then choose two different $\kappa$, implementing weak and strong coupling in this sense. Furthermore, we extract the full-width-at-half-maximum range for each dynamics and denote the corresponding value as relaxation times $\tau_R(\kappa)$. Based on these relaxation times, we determine appropriate driving strengths $\lambda$ according to the following scheme:  {We consider a  Rabi cycle, as it would occur if the spin was driven with some strength $\lambda$ but detached from the bath, and we denote its duration by $T_{\text{Rabi}}(\lambda)$ which is for resonant driving given by $\frac{2\pi}{\lambda}$.} Then we choose for either coupling strength $\kappa$ corresponding driving strengths $\lambda$ such that $T_{\text{Rabi}}(\lambda)\approx \tau_R(\kappa)$. We call the resulting drivings ``weak driving'' and ``strong driving'', respectively. Since we eventually apply both drivings to both couplings, we get four different scenarios for which we proceed by computing the respective four work pdf's. This scheme is intended to identify the most interesting scenarios. To reduce numerical effort, we restrict our investigation to those interesting parameter settings. We expect, however, the main results to apply similarly to almost arbitrary drivings and couplings. None of the calculations done in connection with the work at hand violated this expectation.

\subsection{Density of States of the Spin Ladder}\label{subsec:dos_ladder}
\begin{figure}
\includegraphics[width=1\columnwidth]{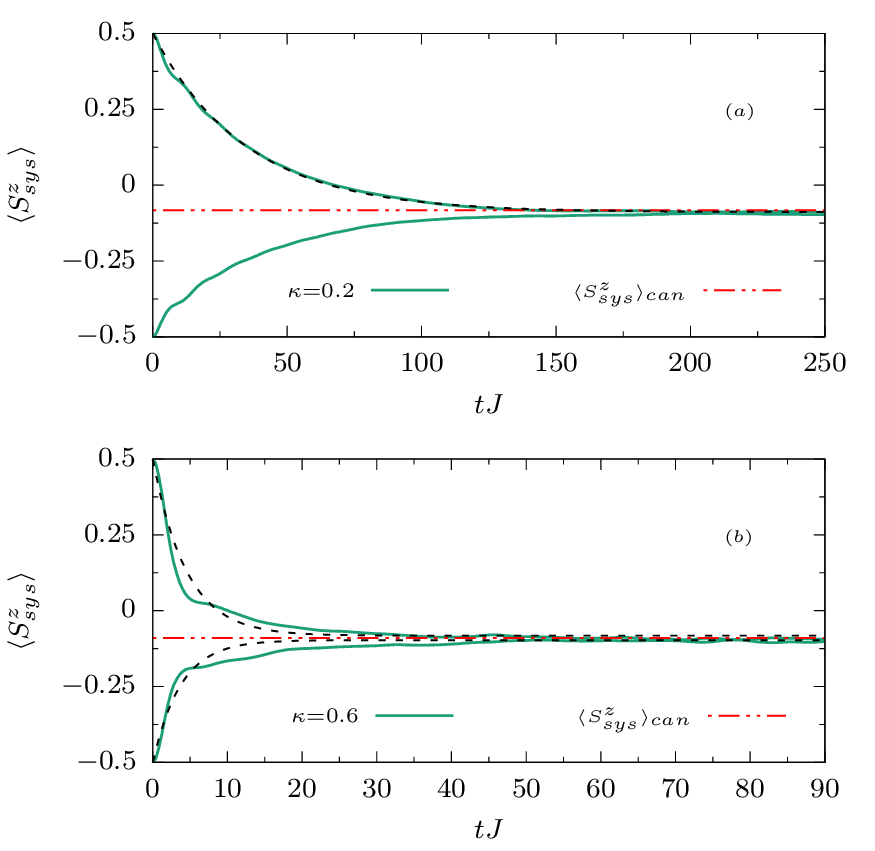}
\caption{Dynamics of the expectation value of the magnetization in $z$ direction, i.e., $\langle S^z_{\text{sys}}\rangle$, for weak ($\kappa=0.2$, upper panel) and strong ($\kappa=0.6$, lower panel) coupling between bath and system for $L=9$. Dashed lines show exponential fits. While for the weak coupling we find exponential decay, we find for strong coupling a strong non-exponential behavior. However, both dynamics tend toward a ``universal'' expectation value corresponding to $\beta\approx0.69$ (see text).}
\label{fig:figure3}
\end{figure}

We start by displaying  in Fig.\ \ref{fig:figure2} semi-log plots of the DOS $\Omega$ for several system sizes $L=6-9$ and system-bath coupling strength $\kappa = 0.2$, within an intermediate energy regime. The energy resolution (graining) is for all cases fix, $\approx0.09$.   {We notice that (i) $\Omega$ becomes smoother with increasing system size as expected due to larger number of energy eigenstates per energy grain, and (ii) the exponential fits (dashed lines with exponent $\beta\approx0.69$) seem to agree on larger energy regimes with $\Omega$ as the system size increases.} Due to these findings, we restrict the following investigations to the energy regimes $[E_0-\Delta/2,E_0+\Delta/2]$\label{p:interval} with $E_0\approx-0.2\,N$ and $\Delta = 2.5$, where $N$ denotes the total number of spins, i.e., $N=2 L + 1$.

\subsection{Relaxation Dynamics of the Spin Coupled to the Spin Ladder}\label{subsec:dyn_ladder}

As mentioned above, we will proceed by briefly discussing the dynamics of the expectation values $\langle S^z_{\text{sys}} (t)\rangle$, i.e., the magnetization in $z$ direction, since  these allow for a classification in terms of  weak and strong system-bath coupling, see Sec. \ref{workdistri}. To this end we define special initial states that are product states of a bath state  ($\pi_{E,\delta}^{\text{bath}}$) featuring a narrow energy distribution  and  system states ($\pi_m$) which may be described as either being spin-up or spin-down, i.e.\, featuring magnetic quantum numbers $m = \pm 1/2$. Hence, their definition reads
\begin{equation}
 \rho = \frac{\pi_{E,\delta}^{\text{bath}} \otimes \pi_m}{\text{Tr}\left\{\pi_{E,\delta}^{\text{bath}} \otimes \pi_m\right\}}~~~,
 \label{eq:state_relaxation}
\end{equation}
where $\pi^{\text{bath}}_{E,\delta}$ denotes an energy projector onto an energy shell of the isolated bath described by $H^\prime_{\text{ladder}} $, with mean energy $E$ and width $\delta$, and $\pi_m$ denotes a projector onto the eigenstates of $S_{\text{sys}}^z$ respectively. Note that we restrict the dynamics to the magnetization subspace corresponding to $M_{\text{tot}}=\sum^N_{i=1}\,S^z_i=(N-1)/2$, i.e., one of the largest magnetization subspaces.  In Fig. \ref{fig:figure3} we summarize the results for $\kappa = 0.2$, which we identify as weak coupling, and $\kappa = 0.6$, which we identify  as strong coupling. As outlined in Sec. \ref{workdistri}, the distinction of weak and strong coupling is reasoned by the types of relaxation behavior that these cases yield, i.e., for $\kappa=0.2$ one finds an 
exponential relaxation (with relaxation time $\tau_R \approx 24$), as expected in the weak coupling limit \cite{Bartsch2007,Niemeyer2013}, whereas for $\kappa=0.6$ the behavior for small times is clearly non-exponential (with relaxation time $\tau_R \approx 2.5$).\\
As outlined in Sec. \ref{workdistri}, we choose, in order to implement weak and strong driving $\lambda = 0.26\, (T_{\text{Rabi}} \approx 24.16)$ and  $\lambda = 2.5\, (T_{\text{Rabi}} \approx 2.51)$. Eventually the duration of the irradiation must be fixed. According to Sec. \ref{sec:workdistri}, we intend to induce work pdf's that essentially live in an exponential region of the DOS, i.e., $[E_0-\Delta/2,E_0+\Delta/2]$. Given this condition, the weak irradiation or driving may be applied longer than the strong driving. As will turn out below, a duration of the driving of 6.5 periods in the case of weak driving and a duration of a half period in the case of strong driving are reasonable choices.

A further comment on the relaxation dynamics itself may be in order. Irrespective of the coupling strengths and the initial value,  $\langle S^z_{\text{sys}} (t)\rangle$ appears to converge against an ``univeral'' expectation value. This value is in accord with the corresponding standard, canonical equilibrium value at inverse temperature $\beta \approx 0.69 $, as suggested by Fig.\ \ref{fig:figure2}. This example thus agrees with the principles of ``canonical typicality'' \cite{Goldstein2006, bartsch2009} and the ETH in the sense discussed in Ref.\  \cite{bartsch17}.

\begin{figure}
\includegraphics[width=1\columnwidth]{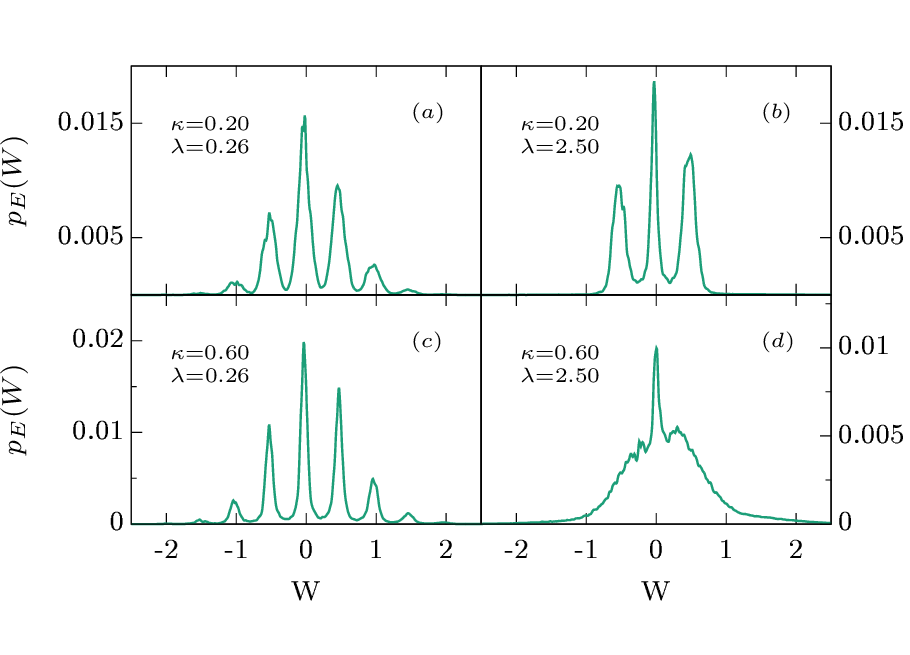}
\caption{Work distributions for $L=8$ and all four combinations of $\kappa =0.2,0.6$ and $\lambda=0.26,2.5$. Note that for $\lambda=0.26$ the duration of the protocol is $T = \frac{13\pi}{\omega}$ whereas for $\lambda=2.5$ it is only $T = \frac{\pi}{\omega}$ (see text).}
\label{fig:figure4}
\end{figure}

\begin{figure}
\includegraphics[width=1\columnwidth]{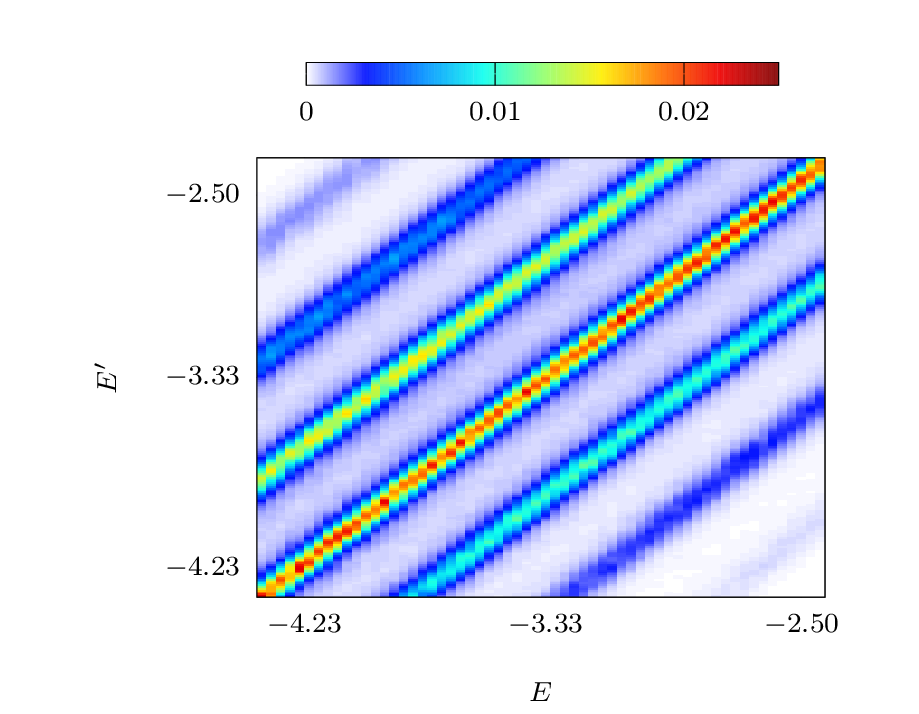}
\caption{Colormap of energy distributions after the driving protocol for severval initial energies, i.e., $p_{E}(E' -E)$ vs. $E$ and $E^\prime$. The parameters are $L=8,\kappa=0.2,\lambda=0.26$.}
\label{fig:figure5}
\end{figure}


\section{Probability Distributions of Work for the Spin-ladder model} \label{sec:work_spin_ladder}

With all parameters set, numerical simulations of the spin-resonance experiment as outlined in Sec.\ \ref{sec:model} can be performed and work pdf's according to Eq.\ (\ref{eq:PEW}) can be extracted. Some exemplary results (for $L = 8 , E \approx -3.2$) are displayed in Fig.\ \ref{fig:figure4}. The work pdf's for weak driving are qualitatively nicely interpretable in terms of time-dependent perturbation theory: Since the spin is resonantly exposed to radiation of frequency $\nu = 0.5$, energy may be absorbed or released in quanta of $\epsilon = 0.5$. This corresponds to the peaks appearing in distances of $\Delta W = 0.5$. If the irradiation endures, a second, third, etc.\  quantum of amount $\epsilon$ may be absorbed/released. Thus, more peaks appear at the sides, with decreasing intensity, though. The only work pdf that defies this interpretation is the one corresponding to strong interaction and strong driving (lower right panel). Since the interpretation roots in time-dependent perturbation theory, this is hardly surprising.\\ 

\begin{figure}
\includegraphics[width=1\columnwidth]{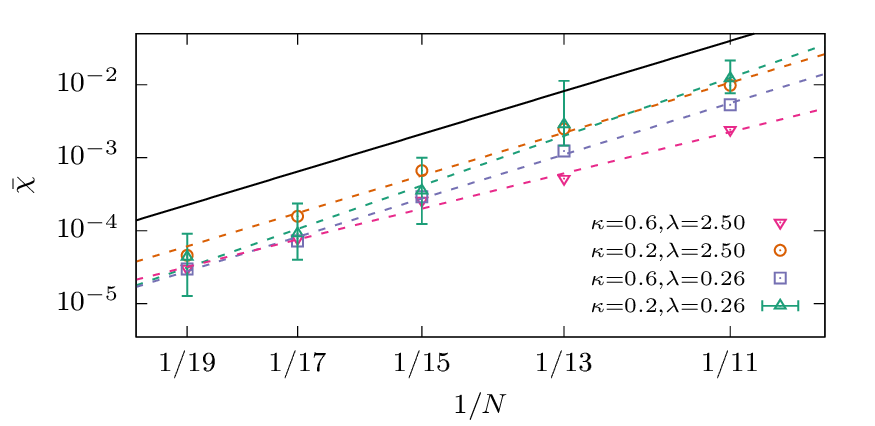}
\caption{ {Quantifying the differences between work distributions corresponding to different initial energies and the parameters $\lambda,\kappa$ discussed above. (Dashed lines indicate power-law fits and, as guide to the eye, the solid line indicates a power-law $\propto N^{-\alpha}$ with mean exponent w.r.t. these fits, i.e., $\alpha\approx9.5$.) For clarity reasons we restrict the display of the smallest and largest value of $\chi(E_0,E^\prime)$ within the consider energy regime to the data corresponding to $\kappa=0.2,\lambda=0.26$ (vertical bars). For the other cases there is a similar behavior.}}
\label{fig:figure6}
\end{figure}

We now turn to the central question of the probability distributions being stiff. Fig.\ \ref{fig:figure5} shows a colormap of energy distributions after the driving protocol, i.e., $p_{E}(E' -E)$, for initial energies $E$ from the energy interval on the horizontal axis. The parameters are $L=8,\kappa=0.2$ and $\lambda = 0.26$. In this display style, a structure which is invariant w.r.t. translations along the diagonal of the graph indicates a stiff work pdf in the sense of Eq.\ (\ref{eq:stiff}). Thus qualitatively stiffness may already be inferred from Fig.~\ref{fig:figure5}. In order to investigate stiffness more  thoroughly, we calculate the corresponding  $\bar{\chi}$ according to Eq.\ (\ref{eq:barchi}) for different system sizes.  {The results are displayed in Fig.~\ref{fig:figure6} for $E_0\approx -0.2\,N$ and the energy regime determinated on page \pageref{p:interval}.} These data strongly indicate that in the limit of large systems the work pdf's indeed become strictly stiff, as  $\bar{\chi}$ quickly tends to zero as $1/N\rightarrow0$ $(N\rightarrow\infty)$. 
This appears to be true for all cases considered, i.e., regardless of whether the coupling is weak or strong and regardless of whether the driving is weak or strong. This finding is a main result of the paper at  hand. While we consider it important on its own right, it also implies the validity of the JR for massively non-Gibbsian, e.g.\ , microcanonical initial states. 


\section{preliminary considerations on the spin-chain setup: density of states and free relaxation dynamics}\label{sec:prechain}

Since, as argued in Sec. \ref{sec:workdistri}, stiff pdf's may only occur in regions featuring an exponential DOS, we start (just like in the previous case of the spin ladder) by identifying such exponential regions in the spin chains' DOS. Here we intend to choose the simplest model by setting all coupling strengths to $1$. Thus, the full system is indeed integrable in the sense of a Bethe-ansatz \cite{bethe_r}. For completeness, we compute the free relaxation dynamics also for this model.

\subsection{Density of States of the Spin Chain}\label{subsec:dos_chain}

\begin{figure}
\includegraphics[width=1\columnwidth]{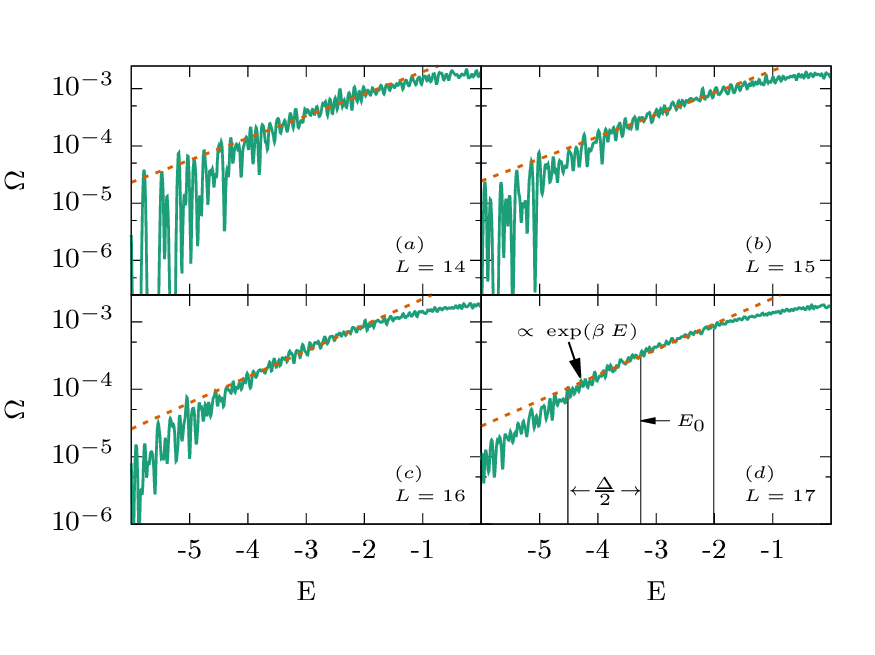}
\caption{Semi-log plots of the density of states for the chain model for $L = 14 - 17$. As in Fig.~\ref{fig:figure2} dashed lines indicate exponential fits.  {Here, also, it seems that the energy range for which $\Omega$ agrees well with the fits increases as the system size rises.}}
\label{fig:figure8}
\end{figure}

As before we start with the DOS for the spin chain. It is presented in Fig.\ \ref{fig:figure8}. In accord with the literature \cite{forster}, the calculations indicate that there are energy regimes in which the DOS is well described by an exponential growth, e.g.\ , at $E_0 \approx -0.18 \, (L+1)$ and $\Delta = 2.5$. \label{regime2} We find the DOS well described by $\beta \approx 0.86$. As for the ladder model, the agreement between $\Omega$ and the exponential fits increases as the system sizes becomes larger. Thus also for this model, regardless of its integrability, stiff work pdf's are possible.

\begin{figure}
\includegraphics[width=1\columnwidth]{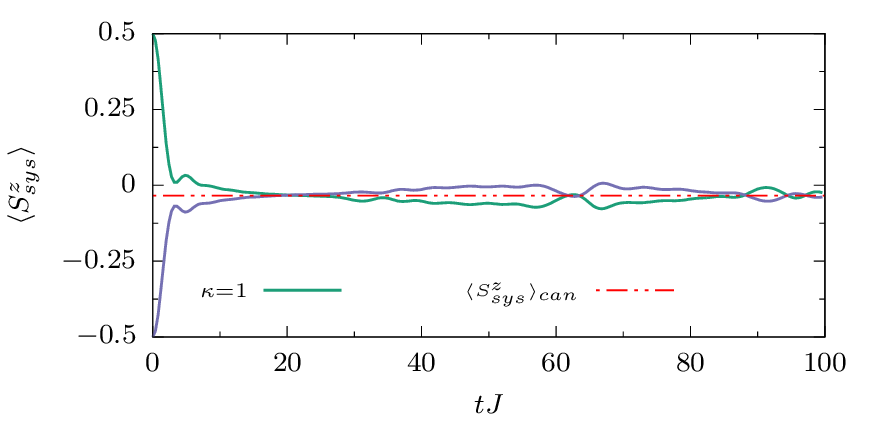}
\caption{Dynamics of the magnetization for the integrable chain model ($L=18$). The relaxation behavior is strongly non-exponential and features fluctuations even for large times. Nevertheless, the dynamics fluctuate around an ``universal'' expectation value corresponding to $\beta\approx0.86$. The relaxation time $\tau_R$ is similar to the one for the strongly coupled ladder system.}
\label{fig:figure9}
\end{figure}


\subsection{Relaxation Dynamics of the Spin Coupled to the Spin Chain}\label{subsec:dyn_chain}

Again we construct an initial state according to Eq.~(\ref{eq:state_relaxation}) with $\pi^{\text{bath}}_{E,\delta}$ based on $H^\prime_{\text{chain}}$ and compute the dynamics for the magnetization subspace $M_{\text{tot}}=(N-1)/2$. The relaxation behavior of $\langle S^z_{\text{sys}} \rangle$ is displayed in Fig.~\ref{fig:figure9}. It appears to be similar to the one for the strongly coupled ladder model. We notice, however, that $\langle S^z_{\text{sys}} \rangle$ fluctuates around some long-time average. Up to the times shown here, apparently these fluctuations are no finite-size effect but persist in the limit of large systems. This principal difference reflects the integrability of the system. Nevertheless, irrespective of the initial value, $\langle S^z_{\text{sys}} (t)\rangle$ appears to converge against, or fluctuate around, an ``universal'' expectation value. This value is also in accord with the corresponding standard, canonical equilibrium value at inverse temperature $\beta \approx 0.86 $, as suggested by Fig.\ \ref{fig:figure8}. Regardless of integrability this  example thus also agrees with the principles of ``canonical typicality'' \cite{Goldstein2006, bartsch2009} and the ETH in the sense discussed in Ref.\  \cite{bartsch17}. For a more detailed discussion of thermalization in the presence of conservation laws, see e.g.\  \cite{sirker}.
 
The relaxation time $\tau_R \approx 1.6 $ is even smaller than for the strongly coupled ladder model. In order to achieve $\tau_R \approx T_{\text{Rabi}}$ we have to use a rather strong driving, i.e., $\lambda\approx3.85$. For the work pdf's to be restricted to the exponential regime of the corresponding DOS, we furthermore choose $\nu=0.75$ where the protocol duration is given by $\pi/\nu$.  {Note that because of the large $\lambda$ and ``small`` $\nu$ detuning corrections \cite{Tannoudji2006,atkins2011} for the Rabi frequency due to the non-resonant driving are negligible here.} 


\section{Probability Distributions of Work for the Spin Chain model} \label{sec:work_spin_chain}
\begin{figure}
\includegraphics[width=1\columnwidth]{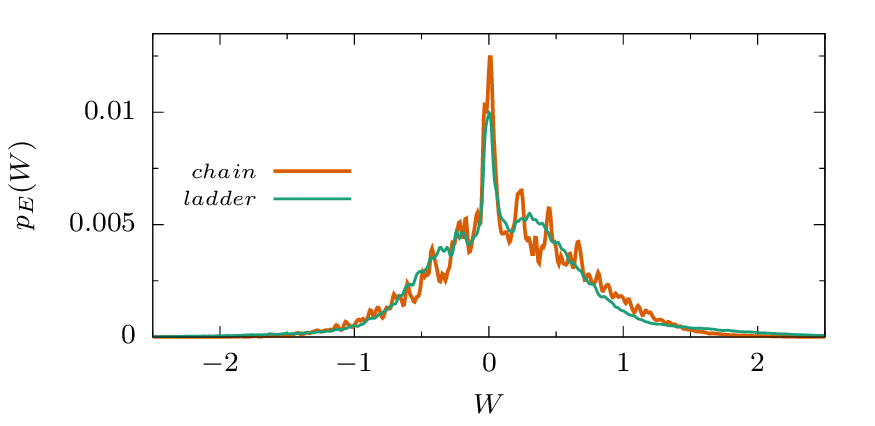}
\caption{Comparison of work pdf's for the strongly driven chain model and the ladder model ($\kappa=0.6,\lambda=2.5$). Apart from minor peaks in the work pdf of the chain model the general shape of both distributions is very similar, although the chain model is integrable and the ladder model is non-integrable; see text for details.}
\label{fig:figure_pro_chain}
\end{figure}

\begin{figure}
\includegraphics[width=1\columnwidth]{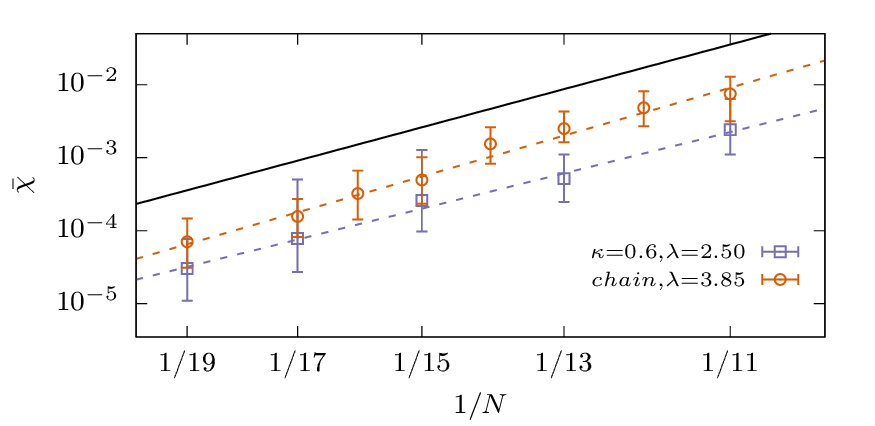}
\caption{Comparison of $\chi$ for the ladder model ($\kappa=0.6,\lambda=2.5$) and the chain model for $\lambda=3.85$. Dashed lines show power law fits while vertical bars indicate the smallest and largest value of $\chi(E_0,E^\prime)$. (As guide to the eye the solid line indicates a power law $\propto N^{-\alpha}$ with mean exponent $\alpha\approx8.4$.) Apparently for the chain model the distributions become initial-energy independent in the thermodynamic limit also.}
\label{fig:figure11}
\end{figure}

As in Sec. \ref{sec:work_spin_ladder}, we again calculate the work pdf's according to the prior identified parameters and display exemplarily in Fig. \ref{fig:figure_pro_chain} for $L=18$ a corresponding work pdf. For comparison, Fig. \ref{fig:figure_pro_chain} also includes a work pdf for the strongly coupled and strongly driven ladder model ($\kappa=0.6,\lambda=2.5$). In order to analyze the stiffness of the work pdf's, we again compute $\bar{\chi}$ for increasing chain lengths and display the results in Fig.\ \ref{fig:figure11} (for better comparability, Fig.\ \ref{fig:figure11} also shows the results for the strongly coupled, strongly driven ladder model, $\kappa=0.6,\lambda=2.5$). The behavior of stiffness of the work pdf's appears to be very similar to the one for the non-integrable ladder model, the data indicate that also for the integrable chain model the work pdf's become strictly stiff in the limit of large system sizes. This finding is another main result of the paper at 
hand. Just like in the case of the ladder model, it implies the validity of the JR also for massively non-Gibbsian, i.e., microcanonical initial states. Thus, evidently, this asymptotic extended validity is not restricted to chaotic models, otherwise it could not occur for a clean spin chain.

\section{Stiffness of Fermi's Golden Rule Rates and Eigenstate Thermalization Hypothesis}\label{fgr}

Since all the various examples addressed so far point to stiff work pdf's in the limit of large systems, the question arises whether this phenomenon can be made accessible by more general concepts.  As a first step in this direction we analyze the structure of the matrices that represent the driving operator ${S}^x_{\text{sys}}$, displayed in the basis formed by the energy eigenstates of the non-driven system. In the case of weak driving, a link between the work pdf's and the above driving matrices arises from Fermi's Golden Rule (FGR). As mentioned before, the work pdf's may simply be viewed as transition probabilities from energy $E$ to energy $E+W$. In the weak-driving case, these transition rates may be well-described by rates as calculated from FGR which we denote by $\gamma_{E \rightarrow E'}$. The latter, however, have to be computed from the driving (perturbation) operator w.r.t. the above basis of energy eigenstates (for details, see below). Thus, if the actual dynamics are well captured by a FGR approach, the stiffness of the work pdf's implies a corresponding stiffness of the FGR rates, namely 
\begin{equation}
\label{eq:stiffrates}
\gamma_{E \rightarrow E'}\stackrel{!}{=} \gamma(E-E') ~~~.
\end{equation}
This can, however, only hold if the matrix representing the driving operator features a corresponding structure.  {(We investigate the structure of the corresponding matrix numerically in Appendix \ref{A-fermi}.)} In general, the FGR rate for the transition from an energy eigenstate $| i \rangle$ from a small interval around $E$ into the set of eigenstates $\{ | f \rangle \}$ that span the energy interval 
around $E'$ under the influence of some (weak) driving of the form $H_D(t)= \sin (|E-E'|\,t)\,{V}$ (i.e.~in our specific case ${V}=\lambda S^x_{\text{sys}}$) is given by \cite{sakurai}
\begin{equation}
\label{eq:fgr0}
 \gamma_{i \rightarrow E'}= 2\pi \,\overline{|V_{iE'}|^2}\, \Omega(E'), \quad   \overline{|V_{iE'}|^2} := \frac{1}{N(E')}\sum_{f \; \in \; E'} |V_{if}|^2~~~,
\end{equation}
where $V_{if}:= \langle i | {V} | f \rangle$. The total number of energy eigenstates in the interval around $E'$ is denoted by $N(E')$ and ``$f \; \in \; E'$'' is short for: ``such that the energy eigenvalue of  $| f \rangle$ lays in the interval around $E'$''. The energy intervals are to be chosen according to the scheme introduced above Eq.\ (\ref{eq:inistat}) and discussed below Eq.\ (\ref{eq:PEW}). This implies, e.g.\ , $N(E')=\delta \cdot \Omega(E')$. In order to convert the transition probability from a single state to an energy interval, $ \gamma_{i \rightarrow E'}$, to the transition probability from an energy interval to another energy interval, $ \gamma_{E \rightarrow E'}$, one has to coarse grain, i.e., to average:  
\begin{equation}
\label{eq:fgr1}
 \gamma_{E \rightarrow E'}=\frac{1}{N(E)} \sum_{i \; \in \; E}  \gamma_{i \rightarrow E'}~~~,
\end{equation}
where the notation is completely analogue to that used in Eq.\ (\ref{eq:fgr0}). Exploiting Eq.~(\ref{eq:fgr0}), a reformulation of Eq.~(\ref{eq:fgr1}) reads:
\begin{equation}
\label{eq:fgr2}
 \gamma_{E \rightarrow E'}= \frac{2\pi }{\delta \cdot  N(E)} \sum_{\substack{i \; \in \; E \\ f \; \in \; E'   }}   |V_{if}|^2
\end{equation}
Whether or not Eq.\ (\ref{eq:stiffrates}) applies obviously depends on the matrix elements $V_{if}$. On one side, this  can simply be checked for specific cases by computing the FGR rates according to Eq.\ (\ref{eq:fgr2}) numerically. (Of course this requires the numerically exact diagonalization of the undriven Hamiltonians $H_{\text{ladder}}^{0},H_{\text{chain}}^{0}$). On the other side, Eq.\ (\ref{eq:fgr2}) can be used to show that the condition  (\ref{eq:stiffrates}) is closely related to quantities occurring in the so-called ETH ansatz. The latter may be described as follows.

Within the framework of the ETH, it has been suggested that the matrix representation of a generic few-body operator ${V}$ w.r.t the basis formed by the eigenstates $|n \rangle ,|m \rangle$ of an Hamiltonian should be in accord with the following description  \cite{Srednicki,ETHrev,Rigol2008}:
\begin{equation}
V_{mn} = \mathcal{V}(\bar{E})\, \delta_{mn} + \Omega \left(\bar{E}\right)^{-\frac{1}{2}}\,f_V\left(\bar{E},\omega\right)\,R_{mn}~~~,
 \label{eq:ETH}
\end{equation}
where $\bar{E} = (E_m+E_n)/2,\,\omega = E_n-E_m$. Furthermore, both, $\mathcal{V}\left(\bar{E}\right)$ and $f_V \left(\bar{E},\omega\right)$, are assumed to be ``smooth functions of their arguments'' \cite{Srednicki} and $R_{mn}$ may be ``conveniently thought of as independent random variables with zero mean and unit variance'' \cite{Srednicki}. For a detailed discussion on the comparability of many-body Hamiltonians and random matrices see \cite{santos}. (To repeat: often and in all our examples, the $R_{mn}$ are real.)  For the purpose of the present investigation we interpret ``smooth'' in the sense of ``approximately constant on the interval $\delta$''. Hence, we may equivalently define $\bar{E}, \omega$ on the basis of the interval labels rather than the individual energy eigenvalues as $\bar{E}:= (E+E')/2, \omega = E-E'$. Plugging the ansatz in Eq.\ (\ref{eq:ETH}) into Eq.\ (\ref{eq:fgr2}) yields
\begin{equation}
\label{eq:fgr3}
 \gamma_{E \rightarrow E'}= \frac{2\pi }{\delta \cdot  N(E)} \sum_{\substack{i \; \in \; E \\ f \; \in \; E'   }}
 \frac{f^2_V\left(\bar{E},\omega\right)\,R^2_{if}}{\Omega \left(\bar{E}\right)}~~~.
\end{equation}
To an accuracy essentially set by the law of large numbers, this may be rewritten as
\begin{equation}
\label{eq:fgr4}
 \gamma_{E \rightarrow E'}\approx \frac{2\pi }{\delta} 
 \frac{f^2_V\left(\bar{E},\omega\right)N(E')}{\Omega \left(\bar{E}\right)}=
 \frac{2\pi f^2_V\left(\bar{E},\omega\right)\Omega (E')}{\Omega \left(\bar{E}\right)}~~~.
\end{equation}
In order to proceed further, we now specialize in exponential DOS's, i.e.\ , $\Omega (U) \propto e^{\beta U}$, as given in and argued for above Eq.\ (\ref{eq:mustexp}). Doing so yields
\begin{equation}
\label{eq:fgr4}
 \gamma_{E \rightarrow E'}=  2\pi f^2_V\left(\bar{E},\omega\right) e^{-\beta \omega}.
\end{equation}
This unveils that the stiffness of the FGR rates (\ref{eq:stiffrates}) depends directly on $ f^2_V\left(\bar{E},\omega\right)$. To be more specific: If $ f^2_V\left(\bar{E},\omega\right)$ varies negligibly with  $\bar{E}$ on an interval $\Delta$ as defined by Eq.\ (\ref{eq:int}), then the  FGR rates are stiff, i.e., Eq.\ (\ref{eq:stiffrates}) applies, since then the r.h.s. of Eq.\ (\ref{eq:fgr4}) only depends on $\omega$. Hence, the approximate independence of  $ f^2_V\left(\bar{E},\omega\right)$ w.r.t. $\bar{E}$ is, together with an exponential DOS and the applicability of FGR, sufficient for the validity of the microcanonical JR at weak driving. This is a central result of the present section.  

\begin{figure}
\includegraphics[width=1\columnwidth]{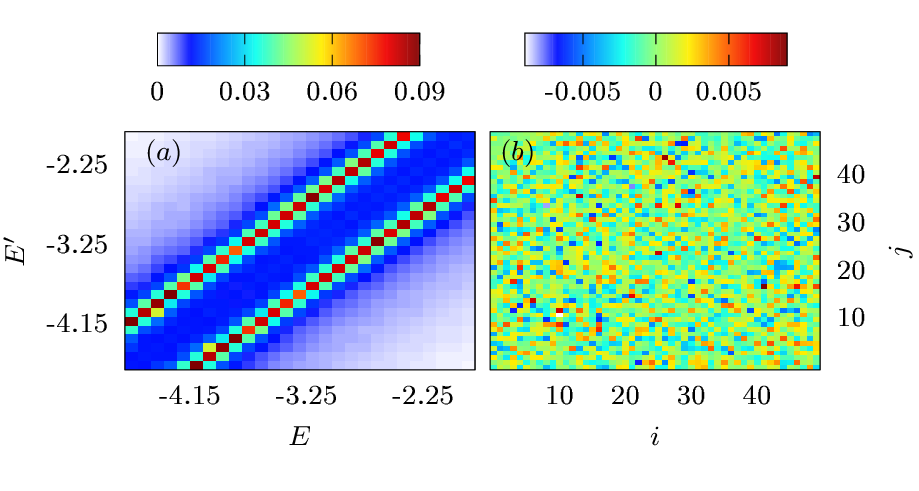}
\caption{(a) Colormap of the coarse grained $S^x_{\text{sys}}$ operator vs. $E,E^\prime$ for the ladder model ($L=7,\kappa=0.2$) for an exemplary energy section. (b) $50 \times 50$ matrix elements of the $S^x_{\text{sys}}$ operator in energy basis of $H^0_{\text{ladder}}$.}
\label{fig:figure13}
\end{figure}

\begin{figure}
\includegraphics[width=1\columnwidth]{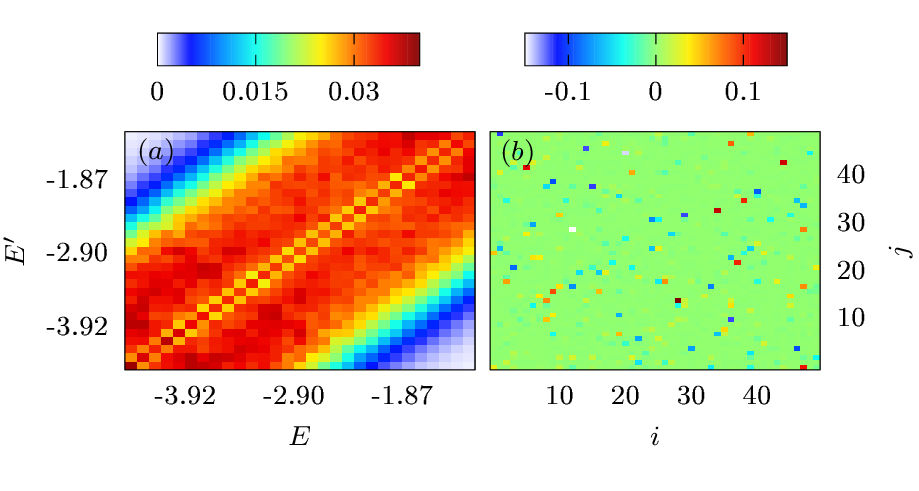}
\caption{(a) Colormap of the coarse grained $S^x_{\text{sys}}$ operator vs. $E,E^\prime$ for the chain model ($L=7$) for an exemplary energy section. (b) $50 \times 50$ matrix elements of the $S^x_{\text{sys}}$ operator in energy basis of $H^0_{\text{chain}}$.}
\label{fig:figure12}
\end{figure}

Motivated by the previous considerations, we proceed by numerically investigating the applicability of the ETH ansatz (\ref{eq:ETH}) to the driving operator $S^x_{\text{sys}}$ for the ladder and the chain scenario. Particularly relevant is the function $ f_V\left(\bar{E},\omega\right)$ and its dependence on $\bar{E}$. In order to visualize $ f_V\left(\bar{E},\omega\right)$, we compute $g(E^\prime,E)$ which is defined as  
\begin{equation}
g(E^\prime,E) := \frac{\sum_{\substack{i \; \in \; E \\ f \; \in \; E'   }}  |V_{if}| \sqrt{N_{(E+E')/2}}}{N_{E^\prime} N_E}~~~.
 \label{eq:ETH3}
\end{equation}
Comparing Eq.\ (\ref{eq:ETH3}) to Eq.\ (\ref{eq:ETH}) reveals that, for $E \neq E'$, 
\begin{equation}
 g(E^\prime,E) \propto |f_V(E',E)|~~~. 
\end{equation}
(Note that, unlike $f_V(E',E)$, $g(E^\prime,E)$ is not independent of $\delta$ on account of the averaging over energy regimes $E,E^\prime$.) In Fig. \ref{fig:figure13} (left panel) we display $g(E^\prime,E)$ for the weakly coupled ladder setup and ${V} \propto S^x_{\text{sys}}$. As varying $\bar{E}$ corresponds to moving parallel to the diagonal on the panel, it is obvious that $g(E^\prime,E)$ is indeed practically independent of $\bar{E}$. (Furthermore, Fig.\ \ref{fig:figure13}(a) essentially reflects the weakly damped Larmor oscillations of the spin's $x$ component). The right panel of Fig. \ref{fig:figure13} visualizes a small sector of the $S^x_{\text{sys}}$ matrix on the level of individual matrix elements. Their values are plotted in false color against their indices. Without any further analysis it appears reasonable to interpret the individual matrix elements as independent random numbers in the sense of the $R_{mn}$ from Eq.~(\ref{eq:ETH}). Performing the same analysis for the integrable chain setup produces data as displayed in  Fig.\ \ref{fig:figure12}. Again, the left panel visualizes $g(E^\prime,E)$ whereas the right panel shows individual matrix elements. Obviously, regardless of integrability, $g(E^\prime,E)$ appears to be independent of  $\bar{E}$ to very good approximation, too. (Moreover, Fig. \ref{fig:figure12}(a) essentially indicates the absence of any Larmor precession in this case and the comparatively quick decay of the   $S^x_{\text{sys}}$ autocorrelation function.) This, in the sense described above, explains the validity of the microcanonical JR in spite of integrability. On the level of individual matrix elements, however, the consequences of integrability are clearly visible: most matrix elements are strictly zero. Since the $S^x_{\text{sys}}$ operator may only couple eigenstates from the same conserved subspaces, the many conservation laws of the spin chain render most matrix elements zero. However, if strictly ordered according to energy, as done here, the non-zero matrix elements appear at more or less random positions. Furthermore their values appear to be random with zero mean. The results on the microcanonical JR suggest that these properties suffice for the applicability of FGR. 

\section{Summary and conclusion}\label{conclusion}
In the paper at hand we investigated the question whether for closed systems (into which a bath may be included), that are driven and initially in a microcanonical state, the resulting work pdf's are independent of the actual initial energy. This assumption, however, denoted as stiffness, can only hold if the DOS of the underlying system is strictly exponential w.r.t. energy. Furthermore, this assumption has far-reaching consequences, e.g.\ , the validity of the Jarzynski relation for a large class of non-Gibbsian states. We studied the stiffness of work pdf's numerically for integrable and non-integrable spin models for various parameter sets. We found that (i) integrability seems to have no influence on the behavior of the stiffness of work pdf's and (ii) for all studied cases stiff work pdf's are expected in the thermodynamic limit. At last we provided a study of the driving operator w.r.t. to Fermi's Golden Rule which may ultimately revealed a relation of our findings to the applicability of the eigenstate thermalization hypothesis to the driving operator.\\

Lastly, we remark the connection of the present findings with the theory of ensemble equivalence in equilibrium statistical mechanics. It is known that for a very large system with an exponential DOS, equilibrium microcanonical and canonical expectations (with a corresponding temperature) of ordinary quantum observables coincide. Here we have been focussing
rather on non-ordinary two-point quantum observable (i.e. the work $W$), and on its statistics under a non-equilibrium setting. 
According, a non equilibrium object, namely the stiffness of the work pdf gives information about the DOS, and in turn about an equilibrium property, namely ensemble equivalence. An interesting question that raises here is whether and to what extent, the microcanonical-work pdf itself coincides with its canonical counterpart, under the assumption of stiffness.


\acknowledgments
We sincerely thank the ``Platform for Fundamental Aspects of Statistical Physics and Thermodynamic`` (Bielefeld, J\"ulich, Oldenburg, Osnabr\"uck) for very fruitful discussions.

\appendix

\section{Energy graining of final work pdf's}\label{A-chi_sec}

\begin{figure}
\includegraphics[width=1\columnwidth]{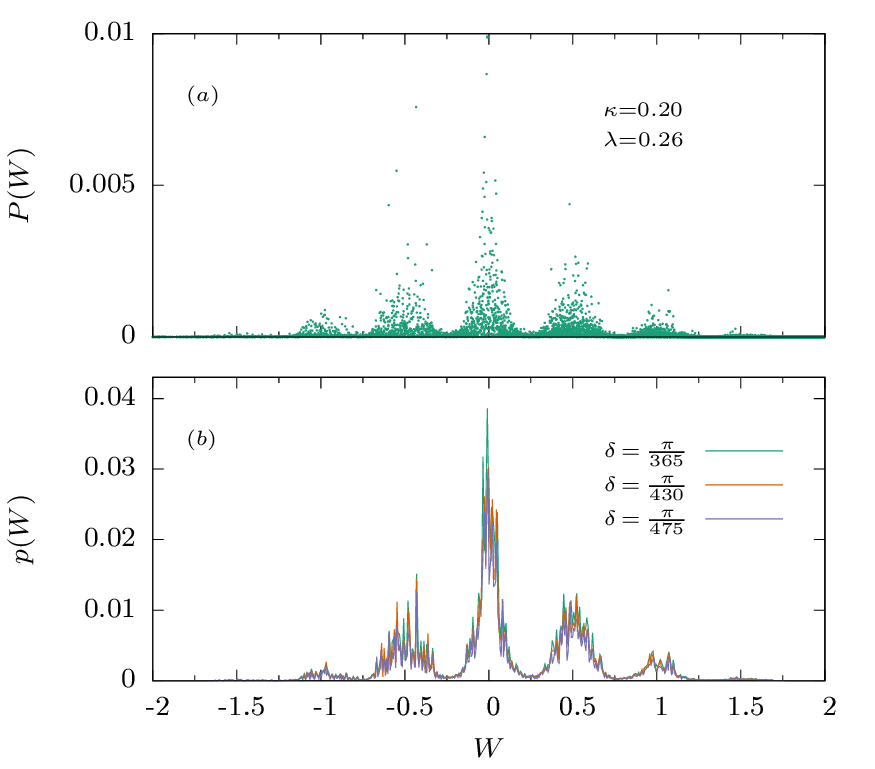}
\caption{(a) Discrete ``ungrained`` work distribution as calculated by Eq.\ (\ref{eq:exactPE}) and (b) ``grained`` work distributions for three distinct graining parameter $\delta$ for $N=15$. All data for the ladder model. The grained versions suggest that the general behavior of the distribution becomes nearly independent of the actual choice of $\delta$.}
\label{fig:figureS1}
\end{figure}

 The exact work distribution $P_{E^\prime}(W)$ after the driving protocol for an initial state starting at energy $E^\prime$ is given by
  \begin{equation}
P_{E^\prime}(W)=\sum_n\,|\langle E_n|{\bf U}(T)| E^\prime\rangle|^2\,\delta(W+E^\prime-E_n)~~~.
 \label{eq:exactPE}
\end{equation}
 For any finite system the energy spectrum is discrete and thus initial states at different energies will in general yield different $P_{E^\prime}(W)$. However, in the thermodynamic limit the energy spectra can be seen as quasi-continuous, i.e., Eq.\ (\ref{eq:exactPE}) may be replaced by
   \begin{equation}
   \begin{split}
P_{E^\prime}(W)&=\int\,dE^{\prime\prime}\,|\langle E^{\prime\prime}|{\bf U}(T)| E^\prime\rangle|^2\,\delta(W+E^\prime-E^{\prime\prime})\\
&= |\langle E^\prime+W|{\bf U}(T)| E^\prime\rangle|^2=c(E^\prime,W)~~~.\\
 \end{split} \label{eq:exactPE_int}
\end{equation}
Since $c(E^\prime,W)$ is per definition continuous and normalized, we can interpret it as a work pdf, i.e., $p_{E^\prime}(W):=c(E^\prime,W)$ as used in the main text. Note that in all cases studied, the energy spectra are far off from being (quasi-)continuous. Nevertheless, it is possible, by properly graining the resulting work distributions as calculated from Eq.\ (\ref{eq:exactPE}), to obtain distributions that behave like actual work pdf's, i.e., are independent of the graining parameter. To do so, we have to make sure that (i) the graining width $\delta$ (or energy resolution) is much smaller than the energy scale on which the details we are interested in appear and (ii) $\delta$ has to be larger than the individual level spacing of the energy spectrum. This gives us a lower and upper bound for $\delta$ where the lower bound decreases with increasing system size due to the exponential growth of the Hilbert space. As prior said, the main indication for a ``good`` graining parameter $\delta$ is given by the fact that the resulting work distribution will be mostly independent of the actual used $\delta$. In order to demonstrate that this is indeed the case for the systems studied, we display for the ladder model ($N=15$) in Fig. \ref{fig:figureS1}(a) the ``ungrained'' work distribution according to Eq.\ (\ref{eq:exactPE}) and in Fig. \ref{fig:figureS1}(b) exemplarily three grained work distributions for several distinct $\delta$.
The lower panel clearly suggests that the general shape of the work distribution is indeed independent of the actual choice of the graining parameter $\delta$. Thus, we are able to interpret these distributions as accurate work pdf's. Note that fluctuation decrease as the system size increases as is the case in e.g.~Fig.~\ref{fig:figure4}.

\section{Solving the time-dependent Schr\"odinger equation (TDSE)}\label{A-tdse}
In the paper at hand we deal with time-dependent Hamiltonians, i.e., $H_{l/c}(t) = H^0_{l/c}+H_D(t,\nu)$ [cf.\  Eq.\ (\ref{eq:ham_time}) in main text], where $H^0_{l/c}$ is the time-independent Hamiltonian that describes the whole system at $t=0$, and $H_D(t,\nu)$ denotes the time-dependent driving operator [see main text Eq.\ (\ref{eq:driving})]. Since we do not need to distinguish between ladder and chain models as is necessary in the main text we omit hereafter the subscripts. Thus, the dynamics are governed by the TDSE
 \begin{equation}
  i\,\frac{\partial}{\partial\,t}|\psi(t)\rangle = H(t)\,|\psi(t)\rangle~~~,
  \label{eq:tdse}
 \end{equation}
 where $|\psi(t)$ denotes the wave function of the system and solutions may be written as
 \begin{eqnarray}
 |\psi(t+\delta t)\rangle = {\bf U}(t+\delta t,t)\,|\psi(t)\rangle\notag~~~,\\
 {\bf U}(t+\delta t,t) = \mathcal{T} \exp\left(-i\,\int^{t+\delta t}_t H(t') dt'\right) ~~.
 \label{eq:itsolv}
 \end{eqnarray}
 Note that we set $\hbar=1$ throughout the entire work.\\
In general, getting these solutions is accompanied by extensive amounts of numerical efforts; see e.g.\  Refs.\cite{deraedt1987,Jin2016}. Here we employ an iterative approximation scheme for small time steps $\delta t$.\\
To do so, we assume that for some sufficiently small $\delta t$ the Hamiltonian can be considered as being constant during each time step, i.e., $H(t) \approx H(t+\delta t/2)$ on the interval $[t,t+\delta t]$. Then the iterative solutions for Eq.\ (\ref{eq:tdse}) read
 \begin{equation}
 |\psi(t+\delta t)\rangle = \exp\left[-i\,H(t+\delta t/2)\,\delta t\right]\,|\psi(t)\rangle~~~.
 \end{equation}
 Hence, to solve the TDSE one can conveniently use any iteration scheme as long as the Hamiltonian fulfills the step-wise-constant condition.\\
 Here we use a fourth-order Runge-Kutta iteration scheme \cite{Elsayed2013,Khodja2016} where in each time step we recalculate the Hamiltonian, i.e., evaluate the present protocol terms $H_D(t+\delta t/2)$. In the following we sketch the general numerical process.
\begin{enumerate}
 \item Calculate the initial Hamiltonian and the initial state, and store them into temporary objects, i.e., ${H^\prime}=H(t=0)$ and $|\phi\rangle=|\psi(t=0)\rangle$
 \item Evaluate the four temporary Runge-Kutta vectors 
 \begin{equation}
  \begin{split}
  |\nu_1\rangle &= i\,\delta t\,H^\prime\,|\phi\rangle~~~~~~~,\\
  |\nu_2\rangle &= i\,\delta t/2\,H^\prime\,|\nu_1\rangle~~~,\\
  |\nu_3\rangle &= i\,\delta t/3\,H^\prime\,|\nu_2\rangle~~~,\\
  |\nu_4\rangle &= i\,\delta t/4\,H^\prime\,|\nu_3\rangle
  \end{split}
 \end{equation}
 \item Collect all Runge-Kutta vectors to form the final state $|\psi(t+\delta t)\rangle = |\phi\rangle+|\nu_1\rangle+|\nu_2\rangle+|\nu_3\rangle+|\nu_4\rangle$
 \item Recalculate $H^\prime$ and $|\phi\rangle$, i.e., $H^\prime=H(t+\delta t)$ and $|\phi\rangle=|\psi(t+\delta t)\rangle$
 \item Repeat 2-4 till the end of the protocol duration
\end{enumerate}
 
For sufficiently small $\delta t$ the accumulated error remains negligible; see e.g.\  \cite{Feiguin2005}.
 
 
\section{Applicability of microreversibility to the models}\label{A-deri4}

As mentioned in the main text, Eq.\ (\ref{eq:mcft}) holds if the Hamiltonian of the system is microreversible. This requirement is sufficient, but not necessary. In this section we assume that $H(t)$ is a real, symmetric matrix for all times $t$ regarding a basis $\mathfrak{B}$, which will be proven to be another sufficient condition in a way similar to microreversibility:
\begin{equation} \label{real_H}
H(t) \equiv H(t)^*
\end{equation}
The matrices in this and in the subsequent equations, which depend on the concrete choice of a basis, are all meant regarding the basis $\mathfrak{B}$. \\
Before describing the details of the proof, we discuss briefly the ladder model again. This model is a spin system with a static magnetic field acting on a single spin of this system. This magnetic field breaks the microreversibility, cf.\  e.g.\  Ref.\ \cite{sakurai}. Nevertheless, the time-dependent Hamiltonian has a real, symmetric matrix representation with respect to the eigenbasis of the $S^z_{\text{sys}}$-operators of the spins (possible choice for $\mathfrak{B}$ in this case). This can be easily seen by expressing the $S^x_{\text{sys}}$ and $S^y_{\text{sys}}$ by the respective creation and annihilation operators.\\
Next we show that these conditions suffice to proof the validity of Eq.\ (\ref{eq:mcft}) even so the system is not microreversible. To this end we start from the left side of Eq.\ (\ref{eq:mcft}). By using Eq.\ (\ref{eq:inistat}) and (\ref{eq:PEW}) we find:
\begin{equation} \label{exact_start}
\begin{split}
& \frac{p_E(W)}{\tilde{p}_{E+W}(-W)} =
	\frac{\Tr(\pi_{E+W,\delta}^f)}{\Tr(\pi_{E, \delta}^i)} Q ~~~,\\
	& Q = \frac{\Tr(\pi_{E+W,\delta}^f U \pi_{E, \delta}^i U^\dagger)}{\Tr(\pi_{E+W,\delta}^f \tilde{U}^{\dagger} \pi_{E, \delta}^i \tilde U)}
\end{split}
\end{equation}
In order to prove that $Q$ is identical to $1$ and thus Eq.\ (\ref{exact_start}) becomes equivalent to Eq.\ (\ref{eq:mcft}), we complex-conjugate the numerator of $Q$, which does not affect its value:
\begin{equation}
	Q = \frac{\Tr(\pi_{E+W,\delta}^f U^* \pi_{E, \delta}^i (U^\dagger)^*)}{\Tr(\pi_{E+W,\delta}^f \tilde{U}^{\dagger} \pi_{E, \delta}^i \tilde U)}
\end{equation}
We used Eq.\ (\ref{real_H}), which implies that $\pi_{E+W, \delta}^f$ and $\pi_{E,\delta}^i$ have real representations regarding $\mathfrak{B}$ as well. In order to show that $Q$ is identical to $1$, we show that
\begin{equation} \label{eqn_U}
 U^* = \tilde{U} ^ \dagger~~~.
\end{equation}
We start from a more explicit form of the time-ordered exponential \cite{zagoskin}:
\begin{equation}
\begin{split}
\tilde U^\dagger = & \sum_{n=0}^\infty \frac{1}{n!} \tilde W_n^\dagger~~~, \\
\tilde W_n = & \int_{\mathfrak{X}^n} \mathrm{d}^n \tau\; \mathcal{T} \left[ \prod_{j=1}^n (-i) \tilde H(\tau_j) \right] ~~~,\\
\mathfrak{X}^n = & [0, t]^n
\end{split}
\end{equation}
Firstly we calculate $\tilde W_n^\dagger$:
\begin{equation}
\tilde W_n^\dagger = \int_{\mathfrak{X}^n} \mathrm{d}^n \tau\, \tilde{\mathcal{T}} \left[ \prod_{j=1}^n i \tilde H(\tau_j) \right]
\end{equation}
$\tilde{\mathcal{T}}$ is the inverse time-ordering operator. Next we replace $\tilde H$ with its definition and apply the following integral transformation:
\begin{equation}
\phi: \mathbb{R}^n \rightarrow \mathbb{R}^n, \tau_j \mapsto T-\tau_j
\end{equation}
Since $\phi^{-1}(\mathfrak{X}^n) = \mathfrak{X}^n$ and $|\mathrm{det}(\mathrm{d}\phi)| = 1$ the integral finally becomes:
\begin{equation}
\tilde W_n^\dagger = \int_{\mathfrak{X}^n} \mathrm{d}^n \tau\, \mathcal{T} \left[ \prod_{j=1}^n i H(\tau_j) \right] = W_n^*
\end{equation}
$W_n$ is defined in the same way as $\tilde{W}_n$, just for $H$ instead of $\tilde H$. In the last step we used Eq.\ (\ref{real_H}) again. Thus, Eq.\ (\ref{eqn_U}) is valid where the necessity of microreversibility is replaced by the validity of Eq.\ (\ref{real_H}).\\

Generally, the fluctuation theorem holds whenever the Hamiltonian $H(t)$ is invariant under the action of any anti-unitary transformation $K$, at each time $t$. Combination of the property of invariance $[H(t),K]=0$, and anti-unitarity of $K$ (specifically the property $KuK^\dagger=u^*$ for any real $u$ \cite{Messiah62Book}), imply that the instantaneous evolution be reversed by $K$, $K e^{-i H(t) \delta }K^\dagger = e^{iH(t)\delta}$ with $\delta$ an arbitrary small real number. The latter is the crucial property that is employed in proving the fluctuation theorem, either in the microcanonical form of Eq. (\ref{eq:mcft}) or in the canonical form and as well in other forms \cite{Campisi2011,Campisi2014}. Thus invariance under time-reversal (which is a anti-unitary operation) implies the validity of the fluctuation relation \cite{Andrieux2008,Campisi2011}. 
In the case studies presented in the main text, even though a magnetic field is present, which breaks the time-reversal invariance, the Hamiltonian is invariant, at each time $t$, under the complex conjugation $K_\mathfrak{B}$ relative to the representation $\mathfrak{B}$ which is evidently an anti-unitary operator \cite{Messiah62Book}. Accordingly the fluctuation theorem is obeyed.\\
Apart from the findings in the paper at hand, it is worth stressing, that, when a magnetic field is present, $H=H(t,B)$, time reversibility breaks, but since $\Theta\, H(t,B)=H(t,-B)\,\Theta$ holds ($\Theta$:  time reversal operator \cite{Messiah62Book,Campisi2011}), the validity of the FT can be restored under the further provision that the magnetic field be reversed in the backward protocol \cite{Andrieux2008,Campisi2011}.


\section{Generating initial states}\label{A-initialst}
Since for our models exact diagonalization is only feasible for system comprising maximal $15$ spins ($\text{dim}\{\mathcal{H}_{15}\}=32768$), we employ for larger systems a typicality-based approximation scheme \cite{Elsayed2013,Steinigeweg2014,Khodja2016} to mimic the initial states as defined in the main text. To be precise, we generate Gaussian projections of random states $|\phi\rangle$, drawn according to the Haar measure, onto narrow energy shells of the total Hilbert space of the corresponding system, i.e.,
\begin{equation}
  |\psi_E\rangle = C^{-1}\,\,\exp\left(-\frac{(H^0-E)^2}{4\,\sigma^2}\right)\,|\phi\rangle~~~,
  \label{app:is}
\end{equation}
where $C= \langle \phi|\exp\left(-\frac{(H^0-E)^2}{2\,\sigma^2}\right)\,|\phi\rangle$. This state mimics for small $\sigma$ to good accuracy a sharp energy shell around energy $E$. In order to calculate the r.h.s. of Eq.\ (\ref{app:is}) we again employ a fourth-order Runge-Kutta scheme. However, here we replace $H^\prime=(H^0-E)^2$ and use imaginary time steps $\delta t = i\,\delta \tau$. The final ``time`` is correspondingly $T=1/4\,\sigma^2$.\\
Comparison with exact-diagonalization data for $N=15$ yields that $\sigma=1/\sqrt{1000}$ is an appropriate choice to mimic the initial states defined in the main text. Again, we choose such time steps that errors remain negligible.\\
 
 
\begin{figure}
\includegraphics[width=1\columnwidth]{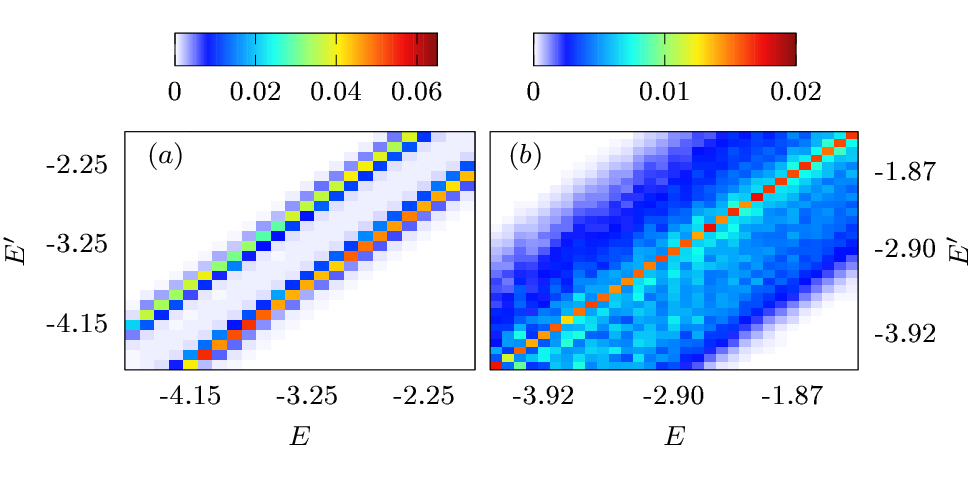}
\caption{Colormap of the energy dependencies of transition rates $\gamma_{E\rightarrow E^\prime}$ vs. $E,E^\prime$ according to Fermi's Golden Rule for the spin ladder (a) and spin chain (b). Both indicate that the transition rates do only depend on the energy difference $E-E^\prime$, i.e., $\gamma_{E\rightarrow E^\prime}=\gamma(E- E^\prime)$, in the energy regimes onto which we focused our investigations.}
\label{fig:S2}
\end{figure}
 
\section{Density of states and energy distributions}\label{A-workpdf}
The density of states (DOS) $\Omega(E)$ can be written as
\begin{equation}
  \Omega(E)=d^{-1}\,\sum_n\,\delta(E-E_n)~~~,
\end{equation}
where $n$ runs over all eigenvalues of $H=H^0(t=0)$ and $d=\text{dim}\{\mathcal{H}\}$ denotes the dimension of the Hilbert space of the respective system. Equivalently, we can express it in terms of time evolutions as
\begin{equation}
  \Omega(E)=\frac{1}{2\pi}\int^{+\infty}_{-\infty}\,e^{i\,E\,t}\,\text{Tr}\left\{e^{-i\,H\,t}\right\}\,dt~~.
\end{equation}
Note that the trace in the latter equation can be evaluated accurately by employing quantum typicality \cite{Hams2000,Elsayed2013,Steinigeweg2014}, i.e.,
\begin{equation}
 \frac{\text{Tr}\left\{e^{-i\,H\,t}\right\}}{d}\approx\langle\phi(0)|e^{-i\,H\,t}|\phi(0)\rangle= \langle\psi(0)|\phi(t)\rangle~~~,
\end{equation}
where $d$ is again the dimension of the underlying Hilbert space and $|\phi(t)\rangle$ is a random state drawn according the Haar measure. The error scales with $1/\sqrt{d}$. As before, $|\phi(t)\rangle$ can conveniently be calculated by a fourth-order Runge-Kutta scheme.\\
Now, the DOS can be approximated by
\begin{equation}
 \Omega(E) \approx C\,\int^{+\Theta}_{-\Theta}\,e^{i\,E\,t}\,\langle\psi(0)|\phi(t)\rangle\,dt~~~,
\end{equation}
where $C$ accounts for the normalization of the DOS and $\Theta$ denotes the time range required to obtain the energy resolution $\pi/\Theta$. Note that the Nyquist sampling theorem presents a restriction according the time steps that can be used (see e.g.\  Ref.\ \cite{Hams2000}). However, here we use such parameters that errors remain small.\\
At last we want to demonstrate that in a very similar way as just presented we can obtain energy distributions of any pure quantum state $|\Psi\rangle$. To do so, we consider the discrete energy distribution
\begin{equation}
 p(E) = \sum_i\,|\langle E_i|\Psi\rangle|^2\,\delta(E-E_n)
\end{equation}
where $|E_n\rangle$ denotes the $i$-th eigenvector. In terms of time evolutions it reads
\begin{equation}
  \begin{split}
 p(E) &= \frac{1}{2\pi}\int^{+\infty}_{-\infty}\,e^{i\,E\,t}\,\langle \Psi|e^{-i\,H\,t}|\Psi\rangle\,dt\\
 &\approx C\,\int^{+\Theta}_{-\Theta}\,e^{i\,E\,t}\,\langle\Psi(0)|\Psi(t)\rangle\,dt~~~.
 \end{split}
\end{equation}
We want to emphasize that the concept of typicality does not enter the last derivation.\\


\section{Investigation of the transition rates according Fermi's Golden Rule}\label{A-fermi}
As discussed in Sec. \ref{fgr} (main text) for weak driving, the internal transition rates from energy $E$ to $E^\prime$ may be well-described by Fermi's Golden Rule. There, we denoted these transition rates as $\gamma_{E\rightarrow E^\prime}$ and the underlying perturbation operator is identified as the $S^x_{\text{sys}}$ operator. Here we will display our results on these $\gamma_{E\rightarrow E^\prime}$ in a similar way as we have done for the ETH-related quantity $g(E^\prime,E)$. In fact, as pointed out in Sec. \ref{fgr}, both quantities may be related to each other. Moreover, if $\Omega(E)\propto e^{\beta\,E}$ holds true, one can calculate the one from the other. Fig.\ \ref{fig:S2} shows the energy dependence of $\gamma_{E\rightarrow E^\prime}$ for the ladder model (left panel) and the chain model (right panel) in the corresponding energy regimes as discussed in the main text. Both matrix presentations indicate that the transition rates do only depend on the energy difference $E-E^\prime$, i.e., $\gamma_{E\rightarrow E^\prime}=\gamma(E- E^\prime)$.

\bibliography{refs}

\end{document}